\documentclass[12pt]{iopart}

\expandafter\let\csname equation*\endcsname\relax 
\expandafter\let\csname endequation*\endcsname\relax 
\usepackage{amsmath}
\usepackage[colorlinks=true,urlcolor=blue]{hyperref}
\usepackage{xcolor}
\begin{document}
\topmargin=-0.5in

\title[Affine Dynamics with Torsion]{Affine Dynamics with Torsion}

\author{Kemal G{\"u}ltekin}

\address{Department of Physics, {\.I}zmir Institute of Technology, 35430 Izmir, Turkey.}
\ead{kemalgultekin@iyte.edu.tr}
\vspace{10pt}
\begin{indented}
\item[]December 2015
\end{indented}

\begin{abstract}
In this study, we give a thorough analysis of a general affine gravity with torsion. After a brief exposition of the affine gravities considered by Eddington and Schr\"{o}dinger, we construct and 
analyze different affine gravities based on the determinants of the Ricci tensor, the torsion tensor,  the Riemann tensor and their combinations. In each case we reduce equations of motion to their simplest 
forms and give a detailed analysis of their solutions. Our analyses lead to the construction of the affine connection in terms of the curvature and torsion tensors. Our solutions of the dynamical equations 
show that the curvature tensors at different points are correlated via non-local, exponential rescaling factors determined by the torsion tensor. 
\end{abstract}

\section{Introduction}
The general theory of relativity is based on the relativistic gravitational action known as the Einstein-Hilbert action \cite{A1, H1}. The metric tensor is the fundamental dynamical variable. With the addition of an appropriate extrinsic curvature the Einstein field equations can be derived. The theory is purely metrical.

Proceeding from Einstein's metric formulation, Eddington considered in 1919 \cite{E3, E1} a reformulation of the theory in terms of only a connection not a metric. His suggestion was altering the metrical gravitational field through the affine connection, which was first realized by Weyl \cite{W1}. Within this context, he proposed a kind of gravitational action constructed by the square root of the determinant of the symmetric Ricci tensor:

\begin{equation}
I_{Edd}=\int d^{4}x \sqrt{\vert\mathcal{R}\vert}\label{1}
\end{equation}
with $\mathcal{R}$ being determinant of $\mathcal{R_{\mu\nu}}$ corresponding to the symmetric part of the Ricci tensor. 
\newline
Without the notion of metric, introducing the affine connection $\Gamma$ as gravitational field and choosing its symmetric one, the variation of the Ricci tensor with respect to that connection is given as follows:
\begin{equation}
\delta {\mathcal{R}}_{\mu\nu} = \nabla_{\rho}\Big( \delta\Gamma^{\rho}_{\mu\nu}\Big) -
\nabla_{\nu}\Big(\delta\Gamma^{\rho}_{\mu\rho}\Big),\label{r} 
\end{equation}
where $\nabla$ is defined as the covariant derivative operator with respect to the affine connection $\Gamma$.
Then, by keeping the proposed action (\ref{1}) stationary, the variation (\ref{r}) leads to the equation of motion:
\begin{equation}
\nabla_{\rho} \left[ \sqrt{{\vert\mathcal{R}}\vert}
\left({\mathcal{R}}^{-1}\right)^{\mu\nu}\right]=0.\label{3}
\end{equation}
A solution to the last equation is provided by defining an invertible tensor field $g_{\mu\nu}$ such that
\begin{equation}
\sqrt{{\vert\mathcal{R}}\vert}
\left({\mathcal{R}}^{-1}\right)^{\mu\nu}=\Lambda \sqrt{{\vert g\vert}}\left(g^{-1}\right)^{\mu\nu},\label{4}
\end{equation}
where $g$ is the determinant of $g_{\mu\nu}$ and $\Lambda$ is a constant. Then Eq.(\ref{4}) ends up in the same form as the Einstein field equations with a purely cosmological constant $\Lambda$:
\begin{equation}
\mathcal{R_{\mu\nu}}=\Lambda g_{\mu\nu},\label{5}
\end{equation}
which shows the significance of the Eddington approach to the theory of gravity by giving a dynamical origin to the general relativity. Moreover, having the equation of motion (\ref{3}) and its solution (\ref{4}) directly result in the compatibility condition $\nabla_{\rho}g_{\mu\nu}=0$ for our metric solution in Eq.(\ref{5}), which shows the crucial result that the symmetric affine connection $\Gamma^{\rho}_{\mu\nu}$ used in Eddington gravity turns out to be the Levi-Civita connection $\lbrace^{\rho}_{\mu\nu}\rbrace_{g}$ itself used in general relativity: 
\begin{equation}
\lbrace^{\rho}_{\mu\nu}\rbrace_{g}=\frac{1}{2}\left(g^{-1}\right)^{\rho\sigma}\left[\partial_{\mu}g_{\sigma\nu}+\partial_{\nu}g_{\sigma\mu}-\partial_{\sigma}g_{\mu\nu}\right].
\end{equation}
Having obtained the Einstein field equations only in vacuum (\ref{5}) without its matter part, which is described by the energy-mometum tensor T$_{\mu\nu}$, is the essential problem of this theory, thus the Eddington gravity is not a complete theory.

To include matter into Eddington's affine gravity, ``Eddington-inspired Born-Infeld gravity''  was proposed with its metric-affine formulation in which both metric and connection fields are taken into account independently \cite{M1}. An Alternative formulation of purely affine gravity in the presence of matter fields was also formulated in \cite{J1}. Recently, another attempt for incorporating matter into the field equations in Eddington's affine picture was provided by ``Riemann-improved Eddington theory'' \cite{D2}. In this study, the Eddington action (\ref{1}) was extended by Riemann curvature so that it led to a dynamical derivation for the complete Einstein field equations. Another recent work was proposed by constructing torsional metric which after using a given Lagrangian density with curvature led to the matter coupling to the affine gravity \cite{P3}. It was also shown that matter can be incorporated when Eddington gravity is formulated in a spacetime that is immersed in a larger eight dimensional space \cite{H2, H3}. As well as making use of an affine framework to include matter into gravity as clarified above, it should also be noticed that the affine framework is particularly useful for addressing the cosmological constant problem properly \cite{D2, H2}. 

Eddington gravity is based on the symmetric (torsionless) affine connection field and on the symmetric Ricci tensor of that connection. In addition to Eddington's approach, a more generalized formulation was given by Schr\"{o}dinger such that he used a nonsymmetric (torsionful) connection field and introduced a nonsymmetric metric tensor \cite{E3}. Unlike Eddington gravity, this metric structure was not obtained dynamically; it was considered as just a prescription. However, using the nonsymmetric metric structure,  which was suggested earlier by Einstein and Straus \cite{ES1}, resulted in the unification of the electromagnetism and gravitation, in which the electromagnetic field is represented by the antisymmetric part of the metric tensor itself \cite{J2}. Some detailed examinations of the nonsymmetric purely affine gravity can also be found in \cite{P3, N1, N2, Shapiro:2001rz}.

In the present work, we give a detailed study of the torsion effects on the purely affine gravity, where we propose torsional action models based on the Ricci and Riemann curvatures as well as the torsion tensor. In our study, the metric structure is not obtained dynamically, then we exclude the metric tensor from our affine formulation. In this respect, we keep the affine connection as the fundamental gravitational field itself.
\newline
In the two torsional actions involving Ricci curvature, we obtain the gravitational field equations for the general, symmetric, and antisymmetric Ricci tensors so that we are able to construct nonsymmetric connection field in the case of the symmetric Ricci tensor and the nonsymmetric contracted connection in the case of an antisymmetric one. The torsionful connection structures together with the torsionful gravitational field equations lead to the non-local, exponential rescaling of the Ricci curvatures considered as general, symmetric, and antisymmetric, and some of the rescalings are determined by the torsion tensor explicitly. Moreover, achieving the affine connection for symmetric Ricci tensor brings about the examination of geodesic equation.  In the action involving the Ricci tensor in which the torsion determinant appears explicitly, we introduce a scalar function including a torsion determinant such that it modifies the results of our action model based on the purely Ricci determinant, except that in both action models for the rescaling of symmetric Ricci tensor we obtain the same form.
\newline
In the other two torsional actions involving the Riemann curvature, we also show the gravitational field equations and for the Riemann curvature in both actions we also define the non-local, exponential rescaling determined by the torsion tensor.  In the action including the Riemann tensor in which the torsion determinant is again seen explicitly, we introduce another scalar function involving the torsion determinant such that it also appears as the modification to the results obtained from the action constructed by the purely Riemann determinant.

Our paper is organized as follows: in Sect. 2, without the notion of metric, we review the Schr\"{o}dinger's generalization of Eddington gravity. In Sect. 3,  we propose the action model, where the torsion determinant appears as an extension to the Ricci curvature. In Sect. 4, inspired by the model given in Sect. 2, we analyze the torsional action based on the Riemann determinant, and in Sect. 5 we examine the last action in which Riemann curvature is modified by the torsion determinant. In Sect. 6, we summarize our results.

\section{Schr\"{o}dinger's generalization of Eddington gravity}

We start the examination by considering an elegant action in which our Lagrangian density is constructed by only the Ricci tensor including the affine connection, such that
\begin{equation}
I_\mathcal{R}=\int d^{4}x\sqrt{\vert\mathcal{R}\vert},\label{7}
\end{equation}
where $\mathcal{R}\equiv \texttt{Det}\left[ \mathcal{R_{\mu\nu}}\right]$ and $\mathcal{R}\equiv \mathcal{R}\left(\Gamma\right)$.
\newline
To see the dynamics of this type of gravitational action (\ref{7}), let us apply the variational principle with respect to the nonsymmetric connection in question as the fundamental gravitational field itself. To this end, the variation of the action is given by
\begin{equation}
\delta I_\mathcal{R}=\int d^{4}x \delta\sqrt{\vert\mathcal{R}\vert}=\frac{1}{2}\int d^{4}x\sqrt{\vert\mathcal{R}\vert}\left(\mathcal{R}^{-1}\right)^{\nu\mu}\delta\mathcal{R_{\mu\nu}}.
\end{equation}
\newline
As we are interested in the variation according to the connection, we take into account the Palatini formula \cite{E1, P1},
\begin{equation}
\delta\mathcal{R_{\mu\nu}}=\nabla_{\rho}\Big( \delta\Gamma^{\rho}_{\mu\nu}\Big)-\nabla_{\nu}\Big( \delta\Gamma^{\rho}_{\mu\rho}\Big)
-2\mathcal{S^{\sigma}\newline_{\rho\nu}}\delta\Gamma^{\rho}_{\mu\sigma}.\label{9}
\end{equation}
In this paper, as we consider the antisymmetric part of the connection, i.e., $\mathcal{S^{\rho}_{\mu\nu}}=\Gamma^{\rho}_{[\mu\nu]}$, as well as the symmetric part of it the last torsional term in the variation (\ref{9}) comes out as the starting point of the torsional contribution to the Eddington gravity, and here we should also notice the tensorial form of the variation of the connection $\delta\Gamma^{\rho}_{\mu\nu}$ \cite{E1}. 
\newline
Then our action becomes
\begin{equation}
\delta I_\mathcal{R}=\frac{1}{2}\int d^{4}x\sqrt{\vert\mathcal{R}\vert}\left(\mathcal{R}^{-1}\right)^{\nu\mu}\left[\nabla_{\rho}\Big( \delta\Gamma^{\rho}_{\mu\nu}\Big)-\nabla_{\nu}\Big( \delta\Gamma^{\rho}_{\mu\rho}\Big)
-2\mathcal{S^{\sigma}\newline_{\rho\nu}}\delta\Gamma^{\rho}_{\mu\sigma}\right].\label{10}
\end{equation}
We improve Eq.(\ref{10}) by the integration by parts for the terms involving covariant derivatives so that we can take advantage of the identity \cite{E1, P2}
\begin{equation}
\int d^{4}x\nabla_{\mu}\Big(\mathcal{J}\mathcal{{V}^\mu}\Big)=2\int d^{4}x\mathcal{S_{\mu}}\mathcal{J}\mathcal{V^{\mu}},\label{11}
\end{equation}
where $\mathcal{J} $ is any scalar density and it corresponds to $\sqrt{\vert\mathcal{R}\vert}$ in our formalism.
\newline
The identity given above is verified by applying the Gaussian theorem to the relation given by combining the covariant derivative of a scalar density   $\mathcal{J}$ \cite{E1},
\begin{equation}
\nabla_{\rho}\mathcal{J}=\partial_{\rho}\mathcal{J}-\Gamma^{\sigma}_{\sigma\rho}\mathcal{J},\label{12}
\end{equation}
with the same derivative of a contravariant vector $\mathcal{V^{\mu}}$ in which process we consider a vanishing hypersurface integral.
\newline
Thus, by taking advantage of Eq.(\ref{11}), the variation under an arbitrary connection becomes
\begin{eqnarray}
\delta I_\mathcal{R}&=\int d^{4}x\lbrace-\frac{1}{2}\nabla_{\rho}\left[\sqrt{\vert\mathcal{R}\vert}\left(\mathcal{R}^{-1}\right)^{\nu\mu}\right]+\frac{1}{2}\nabla_{\sigma}\left[\sqrt{\vert\mathcal{R}\vert}\left(\mathcal{R}^{-1}\right)^{\sigma\mu}\right]\delta_{\rho}^{\nu}\nonumber\\&+\sqrt{\vert\mathcal{R}\vert}\left(\mathcal{R}^{-1}\right)^{\nu\mu}\mathcal{S_{\rho}}-\sqrt{\vert\mathcal{R}\vert}\left(\mathcal{R}^{-1}\right)^{\sigma\mu}\mathcal{S_{\sigma}}\delta_{\rho}^{\nu}\nonumber\\&-\sqrt{\vert\mathcal{R}\vert}\left(\mathcal{R}^{-1}\right)^{\sigma\mu}\mathcal{S^{\nu}\newline_{\rho\sigma}}\rbrace\delta\Gamma^{\rho}_{\mu\nu},\label{13}
\end{eqnarray}
for which the principle of least action, $\delta I_{\mathcal{R}}=0$, results in the most general field equations:
\begin{eqnarray}
\nabla_{\rho}\left[\sqrt{\vert\mathcal{R}\vert}\left(\mathcal{R}^{-1}\right)^{\nu\mu}\right]-\nabla_{\sigma}\left[\sqrt{\vert\mathcal{R}\vert}\left(\mathcal{R}^{-1}\right)^{\sigma\mu}\right]\delta_{\rho}^{\nu}-2\sqrt{\vert\mathcal{R}\vert}\left(\mathcal{R}^{-1}\right)^{\nu\mu}\mathcal{S_{\rho}}\nonumber\\+2\sqrt{\vert\mathcal{R}\vert}\left(\mathcal{R}^{-1}\right)^{\sigma\mu}\mathcal{S_{\sigma}}\delta_{\rho}^{\nu}+2\sqrt{\vert\mathcal{R}\vert}\left(\mathcal{R}^{-1}\right)^{\sigma\mu}\mathcal{S^{\nu}\newline_{\rho\sigma}}=0.\label{14}
\end{eqnarray}
One may go further than Eq.(\ref{14}) by carrying out the contraction with respect to the indices $\mathcal{\rho}$ and $\mathcal{\nu}$ such that it results in
\begin{equation}
\nabla_{\sigma}\left[\sqrt{\vert\mathcal{R}\vert}\left(\mathcal{R}^{-1}\right)^{\sigma\mu}\right]=\frac{4}{3}\sqrt{\vert\mathcal{R}\vert}\left(\mathcal{R}^{-1}\right)^{\sigma\mu}\mathcal{S_{\sigma}},\label{15}
\end{equation}
which, after plugging it into the most general field equations, leads to
\begin{eqnarray}
&\nabla_{\rho}\left[\sqrt{\vert\mathcal{R}\vert}\left(\mathcal{R}^{-1}\right)^{\nu\mu}\right]-2\sqrt{\vert\mathcal{R}\vert}\left(\mathcal{R}^{-1}\right)^{\nu\mu}\mathcal{S_{\rho}}+\frac{2}{3}\sqrt{\vert\mathcal{R}\vert}\left(\mathcal{R}^{-1}\right)^{\sigma\mu}\mathcal{S_{\sigma}}\delta^{\nu}_{\rho}\nonumber\\&+2\sqrt{\vert\mathcal{R}\vert}\left(\mathcal{R}^{-1}\right)^{\sigma\mu}\mathcal{S^{\nu}\newline_{\rho\sigma}}=0.\label{16}
\end{eqnarray}
Here, one more thing to obtain the desired tensorial field equations is to get rid of the scalar density$\sqrt{\vert\mathcal{R}\vert}$. With this aim, multiplying the last equation by $\mathcal{R_{\mu\nu}}$ and implementing to the first term partial differentiation with the fact that 
\begin{equation}
\nabla_{\rho}\sqrt{\vert\mathcal{R}\vert}=-\frac{1}{2}\sqrt{\vert\mathcal{R}\vert}\mathcal{R_{\mu\nu}}\nabla_{\rho}\left[\left(\mathcal{R}^{-1}\right)^{\nu\mu}\right]\label{17}
\end{equation}
 we obtain
\begin{equation}
\nabla_{\rho}\sqrt{\vert\mathcal{R}\vert}=\frac{8}{3}\sqrt{\vert\mathcal{R}\vert}\mathcal{S_{\rho}}.\label{18}
\end{equation}
\newline
After checking, one may realize that Eq.(\ref{17}) is naturally compatible with Eq.(\ref{12}) in which we regard the Jacobi formula that the differential of a determinant \cite{CB}, say $\texttt{Det}\left[\mathcal{A}\right] $ (such as $\texttt{Det}\left[\mathcal{R_{\alpha\beta}}\right]$ in our notation) is equivalent to the trace of the adjoint of a matrix $\mathcal{A}$ multiplied by its differential $d\mathcal{A}$, that is,
\begin{equation}
d\left(\texttt{Det}\left[\mathcal{A}\right]\right)= \texttt{Det}\left[\mathcal{A}\right]\texttt{Tr} \Big[\mathcal{A}^{-1}d\mathcal{A}\Big].\label{19}
\end{equation}
 \newline
Then after using Eq.(\ref{18}) in  Eq.(\ref{16}), we obtain the field equations with the inverse Ricci tensor such that
\begin{equation}
\nabla_{\rho}\left[\left(\mathcal{R}^{-1}\right)^{\nu\mu}\right]+\frac{2}{3}\mathcal{S_{\rho}}\left(\mathcal{R}^{-1}\right)^{\nu\mu}+\frac{2}{3}\mathcal{S_{\sigma}}\delta_{\rho}^{\nu}\left(\mathcal{R}^{-1}\right)^{\sigma\mu}+2\mathcal{S^{\nu}\newline_{\rho\sigma}}\left(\mathcal{R}^{-1}\right)^{\sigma\mu}=0.
\end{equation}\label{20}
After multiplying with $\mathcal{R_{\mu\kappa}}$ and $\mathcal{R_{\xi\nu}}$, the last equation can also be expressed as 
\begin{equation}
\nabla_{\rho}\left[\mathcal{R_{\mu\nu}}\right]-\frac{2}{3}\mathcal{S_{\rho}}\mathcal{R_{\mu\nu}}-\frac{2}{3}\mathcal{S_{\nu}}\mathcal{R_{\mu\rho}}-2\mathcal{S^{\sigma}\newline_{\rho\nu}}\mathcal{R_{\mu\sigma}}=0.\label{21}
\end{equation}

We are now in a position to show significant effects of the tensorial field equations (\ref{21}) on our Ricci tensor, where the situation is now clearly different from Eddington gravity due to the torsion, i.e. the antisymmetric part of the connection, contributions. Let us first see the case by multiplying Eq.(\ref{21}) by the inverse Ricci tensor  $\left(\mathcal{R}^{-1}\right)^{\nu\mu}$ as follows: 
\begin{equation}
\left(\mathcal{R}^{-1}\right)^{\nu\mu}\nabla_{\rho}\left[\mathcal{R_{\mu\nu}}\right]=\frac{16}{3}\mathcal{S_{\rho}},\label{22}
\end{equation}
which can also be seen from Eqs.(\ref{17}) and (\ref{18}). Then we have
\begin{equation}
\left(\mathcal{R}^{-1}\right)^{\nu\mu}\partial_{\rho}\left[\mathcal{R_{\mu\nu}}\right]=\frac{16}{3}\mathcal{S_{\rho}}+2\Gamma^{\beta}_{\beta\rho}.
\end{equation}
Via Eq.(\ref{19}), the last equation leads to
\begin{equation}
\texttt{Det}\big[\mathcal{R_{\mu\nu}}\big(x^{\sigma}\big)\big]=\exp\left\lbrace\int_{x^{\sigma}_{0}}^{x^{\sigma}}\left(\frac{16}{3}\mathcal{S_{\rho}}+2\Gamma^{\beta}_{\beta\rho}\right)dx^{\rho}\right\rbrace\texttt{Det}\big[\mathcal{R_{\mu\nu}}\big(x^{\sigma}_{0}\big)\big]. \label{23}
\end{equation}
Let us now consider the general coordinate transformations $x^{\sigma}_{0}\rightarrow x^{\sigma}(x^{\sigma}_{0})$ under which the Ricci tensor transforms as
\begin{equation}
\mathcal{R_{\mu\nu}}\big(x^{\sigma}\big)=\frac{\partial x^{\alpha}_{0}}{\partial x^{\mu}}\frac{\partial x^{\beta}_{0}}{\partial x^{\nu}}\mathcal{R_{\alpha\beta}}\big(x^{\sigma}_{0}\big). \label{xy}
\end{equation}
Then taking the determinant of Eq.(\ref{xy}) leads to
\begin{equation}
\texttt{Det}\big[\mathcal{R_{\mu\nu}}\big(x^{\sigma}\big)\big]=\left(\texttt{Det}\left[\frac{\partial x^{\mu}_{0}}{\partial x^{\nu}}\right]\right)^{2}\texttt{Det}\big[\mathcal{R_{\mu\nu}}\big(x^{\sigma}_{0}\big)\big]. \label{ba}
\end{equation}
One may here see that the equality of the right-hand sides of Eqs.(\ref{ba}) and (\ref{23}) results in the determinant being given by
\begin{equation}
\texttt{Det}\left[\mathrm{J}^{\mu}_{\nu}\right]=\exp\left\lbrace\int_{x^{\sigma}_{0}}^{x^{\sigma}}\left(\frac{8}{3}\mathcal{S_{\rho}}+\Gamma^{\beta}_{\beta\rho}\right)dx^{\rho}\right\rbrace, \label{ac}
\end{equation}
where we have introduced the Jacobian matrix $\mathrm{J}^{\mu}_{\nu}=\frac{\partial x^{\mu}_{0}}{\partial x^{\nu}}$ responsible for the transformation of the coordinates from $x^{\nu}$ to $x^{\mu}_{0}$. 
\newline
In affine spacetime, a simple relation for the determinant in (\ref{ac}) of the Jacobian can be given by
\begin{equation}
\mathrm{J}^{\mu}_{\nu}=\delta^{\mu}_{\nu}\exp\left\lbrace\int_{x^{\sigma}_{0}}^{x^{\sigma}}\left(\frac{2}{3}\mathcal{S_{\rho}}+\frac{1}{4}\Gamma^{\beta}_{\beta\rho}\right)dx^{\rho}\right\rbrace, \label{cd}
\end{equation}
which has a significant implication on our Ricci tensor; plugging it into Eq.(\ref{xy}) leads to the non-local, exponential rescaling of the Ricci curvature:
\begin{equation}
\mathcal{R_{\mu\nu}}\big(x^{\sigma}\big)=\exp\left\lbrace\int_{x^{\sigma}_{0}}^{x^{\sigma}}\left(\frac{4}{3}\mathcal{S_{\rho}}+\frac{1}{2}\Gamma^{\beta}_{\beta\rho}\right)dx^{\rho}\right\rbrace\mathcal{R_{\mu\nu}}\big(x^{\sigma}_{0}\big).\label{xz}
\end{equation}
The exponential factor
\begin{equation}
\mathcal{K}\left(x^{\sigma},x^{\sigma}_{0}\right)=\exp\left\lbrace\int_{x^{\sigma}_{0}}^{x^{\sigma}}\left(\frac{4}{3}\mathcal{S_{\rho}}+\frac{1}{2}\Gamma^{\beta}_{\beta\rho}\right)dx^{\rho}\right\rbrace
\end{equation}
is here totally responsible for this non-local rescaling such that the torsion vector shifted by the contracted connection provides the Ricci tensor with a conformal mapping from a point $x^{\sigma}_{0}$ to another one $x^{\sigma}$ over spacetime structure including the nonsymmetric affine field \cite{D1}.

Having studied the Ricci tensor with nonsymmetric connection, the field equations (\ref{21}) can now be written in both cases, with symmetric and antisymmetric parts, as \cite{E1}
\begin{equation}
\nabla_{\rho}\mathcal{R_{\left(\mu\nu\right)}}-\frac{2}{3}\mathcal{S_{\rho}}\mathcal{R_{\left(\mu\nu\right)}}-\frac{1}{3}\mathcal{S_{\mu}}\mathcal{R_{\nu\rho}}-\frac{1}{3}\mathcal{S_{\nu}}\mathcal{R_{\mu\rho}}-\mathcal{S^{\sigma}\newline_{\rho\mu}}\mathcal{R_{\nu\sigma}}-\mathcal{S^{\sigma}\newline_{\rho\nu}}\mathcal{R_{\mu\sigma}}=0\label{S1}
\end{equation} 
\begin{equation}
\nabla_{\rho}\mathcal{R_{\left[\mu\nu\right]}}-\frac{2}{3}\mathcal{S_{\rho}}\mathcal{R_{\left[\mu\nu\right]}}+\frac{1}{3}\mathcal{S_{\mu}}\mathcal{R_{\nu\rho}}-\frac{1}{3}\mathcal{S_{\nu}}\mathcal{R_{\mu\rho}}+\mathcal{S^{\sigma}\newline_{\rho\mu}}\mathcal{R_{\nu\sigma}}-\mathcal{S^{\sigma}\newline_{\rho\nu}}\mathcal{R_{\mu\sigma}}=0.\label{S2}
\end{equation}
Here, we have used the indices $\mu$ and $\nu$ as expected from our nonvanishing Ricci determinant in the action (\ref{7}), and it should be noticed that $\mathcal{R_{\mu\nu}}=\mathcal{R_{\left(\mu\nu\right)}}+\mathcal{R_{\left[\mu\nu\right]}}$. 

One may now examine Eq.(\ref{S1}) in the special case where the Ricci tensor is taken to be symmetric, i.e. $\mathcal{R_{\mu\nu}}=\mathcal{R_{\left(\mu\nu\right)}}$. In this case, we get
\begin{equation}
\nabla_{\rho}\mathcal{R}_{\mu\nu}=\frac{2}{3}\mathcal{S_{\rho}}\mathcal{R}_{\mu\nu}+\frac{1}{3}\mathcal{S_{\mu}}\mathcal{R}_{\nu\rho}+\frac{1}{3}\mathcal{S_{\nu}}\mathcal{R}_{\mu\rho}+\mathcal{S^{\sigma}\newline_{\rho\mu}}\mathcal{R}_{\nu\sigma}+\mathcal{S^{\sigma}\newline_{\rho\nu}}\mathcal{R}_{\mu\sigma}.\label{30}
\end{equation}
Owing to the appearance of the symmetric Ricci tensor, the last equation is the best expression so that only it can provide us with a solution for the nonsymmetric connection. Therefore, after the cyclic permutation, Eq.(\ref{30}) gives
\begin{equation}
\nabla_{\mu}\mathcal{R}_{\nu\rho}+\nabla_{\nu}\mathcal{R}_{\rho\mu}-\nabla_{\rho}\mathcal{R}_{\mu\nu}=\frac{2}{3}\mathcal{S_{\mu}}\mathcal{R_{\nu\rho}}+\frac{2}{3}\mathcal{S_{\nu}}\mathcal{R}_{\mu\rho}+2\mathcal{S^{\sigma}\newline_{\mu\rho}}\mathcal{R}_{\nu\sigma}+2\mathcal{S^{\sigma}\newline_{\nu\rho}}\mathcal{R}_{\mu\sigma}.\label{31}
\end{equation}
In addition, using covariant derivatives in Eq.(\ref{31}), we also have
\begin{eqnarray}
\nabla_{\mu}\mathcal{R}_{\nu\rho}+\nabla_{\nu}\mathcal{R}_{\rho\mu}-\nabla_{\rho}\mathcal{R}_{\mu\nu}&=\partial_{\mu}\mathcal{R}_{\nu\rho}+\partial_{\nu}\mathcal{R}_{\rho\mu}-\partial_{\rho}\mathcal{R}_{\mu\nu}\nonumber\\&-2\Gamma^{\sigma}_{\mu\nu}\mathcal{R}_{\rho\sigma}+2\mathcal{S^{\sigma}}\newline_{\mu\nu}\mathcal{R}_{\rho\sigma}\nonumber\\&+2\mathcal{S^{\sigma}}\newline_{\mu\rho}\mathcal{R}_{\sigma\nu}+2\mathcal{S^{\sigma}}\newline_{\nu\rho}\mathcal{R}_{\mu\sigma},\label{32}
\end{eqnarray}
where we have used the definition of the affine connection, i.e. $\Gamma^{\rho}_{\mu\nu}=\Gamma^{\rho}_{\left(\mu\nu\right)}+\Gamma^{\rho}_{\left[\mu\nu\right]}$ with $\Gamma^{\rho}_{\left[\mu\nu\right]}=\mathcal{S^{\rho}\newline_{\mu\nu}}$. Equating the last two equations and multiplying the sides by            
$\left(\mathcal{R}^{-1}\right)^{\alpha\rho}$, we end up with the nonsymmetric connection in the case of the symmetric Ricci tensor:
\begin{equation}
\Gamma^{\rho}_{\mu\nu}=\lbrace^{\rho}_{\mu\nu}\rbrace_{\mathcal{R}_{s}\left(\Gamma\right)}-\frac{1}{3}\left(\delta_{\mu}^{\rho}\mathcal{S_{\nu}}+\delta_{\nu}^{\rho}\mathcal{S_{\mu}}\right)+\mathcal{S^{\rho}\newline_{\mu\nu}},\label{33}
\end{equation}
where the symbol,
\begin{equation}
\lbrace^{\rho}_{\mu\nu}\rbrace_{\mathcal{R}_{s}\left(\Gamma\right)}=\frac{1}{2}\left(\mathcal{R}^{-1}\right)^{\rho\sigma}\left[\partial_{\mu}\mathcal{R}_{\sigma\nu}+\partial_{\nu}\mathcal{R}_{\sigma\mu}-\partial_{\sigma}\mathcal{R}_{\mu\nu}\right],
\end{equation}
is for the notion of a connection known as Christoffel-brackets \cite{E2} in terms of the symmetric Ricci tensor. Thus, the parallel displacement of our vectors from one tangent space to another one over torsionful geodesics is due to the functions of $\mathcal{R}_{\mu\nu}$ and their linear first derivatives in addition to the torsion tensor.
\newline
It is now also possible to search a conformal factor for the symmetric Ricci tensor by multiplying Eq.(\ref{30}) by $\left(\mathcal{R}^{-1}\right)^{\nu\mu}$, 
\begin{equation}
\left(\mathcal{R}^{-1}\right)^{\nu\mu}\nabla_{\rho}\left[\mathcal{R}_{\mu\nu}\right]=\frac{16}{3}\mathcal{S_{\rho}},\label{35}
\end{equation}
which, by considering the general coordinate transformations, leads to 
\begin{equation}
\mathcal{R}_{\mu\nu}\left(x^{\sigma}\right)=\exp\left\lbrace\int_{x_{0}^{\sigma}}^{x^{\sigma}}\left(\frac{4}{3}\mathcal{S_{\rho}}+\frac{1}{2}\Gamma^{\beta}_{\beta\rho}\right)dx^{\rho}\right\rbrace\mathcal{R}_{\mu\nu}\left(x_{0}^{\sigma}\right).\label{36}
\end{equation}
As clearly seen from Eqs.(\ref{35}) and (\ref{36}), and as expected, they have the same form as Eqs.(\ref{22}) and (\ref{23}) except that our Ricci tensor is here symmetric, i.e. $\mathcal{R_{\left[\mu\nu\right]}}=0$; hence we can take advantage of the connection (\ref{33}) by using it in Eq.(\ref{36}). Then the contraction of the connection with respect to $\rho$ and $\mu$ leads to
\begin{equation}
\Gamma^{\beta}_{\beta\rho}=\lbrace^{\beta}_{\beta\rho}\rbrace_{\mathcal{R}_{s}\left(\Gamma\right)}-\frac{8}{3}\mathcal{S_{\rho}},\label{37}
\end{equation}
which after plugging into Eq.(\ref{36}) results in the final expression for the rescaling of the symmetric Ricci tensor:
\begin{equation}
\mathcal{R}_{\mu\nu}\left(x^{\sigma}\right)=\exp\left\lbrace\int_{x_{0}^{\sigma}}^{x^{\sigma}}\frac{1}{2}\lbrace^{\beta}_{\beta\rho}\rbrace_{\mathcal{R}_{s}\left(\Gamma\right)}dx^{\rho}\right\rbrace\mathcal{R}_{\mu\nu}\left(x_{0}^{\sigma}\right),\label{38}
\end{equation}
where $\mathcal{R}_{\mu\nu}$ is therefore rescaled by the non-local, exponential factor given by
\begin{equation}
\mathcal{K}_{s}\left(x^{\sigma},x^{\sigma}_{0}\right)=\exp\left\lbrace\int_{x_{0}^{\sigma}}^{x^{\sigma}}\frac{1}{2}\lbrace^{\beta}_{\beta\rho}\rbrace_{\mathcal{R}_{s}\left(\Gamma\right)}dx^{\rho}\right\rbrace.
\end{equation}
It can also be seen that Eq.(\ref{38}) can be directly obtained from the contracted Christoffel-brackets. 
\newline
Thus, it is essential to emphasize that although we keep the nonsymmetricity of the connection, the torsion effects do not appear explicitly in the conformal mapping of the symmetric Ricci tensor. In other words, the contracted connection (\ref{37}) eliminates the explicit contribution of torsion to the rescaling, and it leaves only the Christoffel-brackets, which have been written in terms of $R_{\mu\nu}(\Gamma)$, as responsible for the conformal transformation.

Additional to the examinations given above, one can decompose the symmetric Ricci tensor under the change of the connection from the Christoffel one such that the separation of the connection can be written as
\begin{equation}
\Gamma^{\rho}_{\mu\nu}=\lbrace^{\rho}_{\mu\nu}\rbrace_{\mathcal{R}_{s}\left(\Gamma\right)}+\mathcal{E}^{\rho}_{\mu\nu},\label{abcd}
\end{equation}
where $\mathcal{E}^{\rho}_{\mu\nu}$ is a tensor due to the difference between the two connections. Then Eq.(\ref{abcd}) leads to the decomposition of the Ricci tensor,
\begin{equation}
\mathcal{R}_{\mu\nu}\left(\Gamma\right)=\mathcal{R}_{\mu\nu}\left(\lbrace^{\rho}_{\mu\nu}\rbrace_{\mathcal{R}_{s}\left(\Gamma\right)}\right)+\nabla_{\rho}\mathcal{E}^{\rho}\newline_{\left(\mu\nu\right)}-\nabla_{(\nu}\mathcal{E}^{\rho}\newline_{\mu)\rho}+\mathcal{E}^{\rho}\newline_{\sigma\rho}\mathcal{E}^{\sigma}\newline_{(\mu\nu)}-\mathcal{E}^{\rho}\newline_{\sigma(\nu}\mathcal{E}^{\sigma}\newline_{\mu)\rho},\label{41}
\end{equation}
where we should be aware of two different Ricci tensors together with two different connection structures such that the first term on the right-hand side is the Ricci tensor $\mathcal{R}_{\mu\nu}\left(\lbrace^{\rho}_{\mu\nu}\rbrace_{\mathcal{R}\left(\Gamma\right)}\right)$ constructed from the connection known as Christoffel-brackets $\lbrace^{\rho}_{\mu\nu}\rbrace_{\mathcal{R}\left(\Gamma\right)}$, which is with respect to the other Ricci tensor $\mathcal{R}_{\mu\nu}\left(\Gamma\right)$ constructed from the affine connection $\Gamma^{\rho}_{\mu\nu}$.The covariant derivatives are here taken with respect to  the Christoffel-brackets. Actually, since  $\mathcal{R}_{\mu\nu}\left(\Gamma\right)$ is repeated in $\mathcal{R}_{\mu\nu}\left(\lbrace^{\rho}_{\mu\nu}\rbrace_{\mathcal{R}\left(\Gamma\right)}\right)$ Eq.(\ref{41}) is not a finite expression, however; it may be good to write the decomposition to see the difference from the torsional action as will be examined after this section. Then by considering the affine connection (\ref{33}), plugging 
\begin{equation}
\mathcal{E}^{\rho}_{\mu\nu}=-\frac{1}{3}\left(\delta_{\mu}^{\rho}\mathcal{S_{\nu}}+\delta_{\nu}^{\rho}\mathcal{S_{\mu}}\right)+\mathcal{S^{\rho}\newline_{\mu\nu}}
\end{equation}
into Eq.(\ref{41}) we end up with the decomposition determined by the torsional terms:
\begin{equation}
\mathcal{R}_{\mu\nu}\left(\Gamma\right)=\mathcal{R}_{\mu\nu}\left(\lbrace^{\rho}_{\mu\nu}\rbrace_{\mathcal{R}\left(\Gamma\right)}\right)-\frac{1}{3}\mathcal{S}_{\mu}\mathcal{S}_{\nu}-\mathcal{S}^{\rho}\newline_{\mu\sigma}\mathcal{S}^{\sigma}\newline_{\nu\rho}.\label{43}
\end{equation}

As a final study for the symmetric Ricci tensor, let us consider the geodesic equation in curved spacetime such that it is produced by the condition defined by the parallel displacement of the tangent vector $\frac{dx^{\mu}}{d\lambda}$ over a path $x^{\mu}(\lambda)$ \cite{C1}:
\begin{equation}
\frac{dx^{\sigma}}{d\lambda}\nabla_{\sigma}\frac{dx^{\mu}}{d\lambda}=0.
\end{equation}
With the condition given above the geodesic equation is then formed into
\begin{equation}
\frac{d^{2}x^{\rho}}{ds^{2}}+\Gamma^{\rho}_{\mu\nu}\frac{dx^{\mu}}{ds}\frac{dx^{\nu}}{ds}=0,\label{45}
\end{equation}
by which it is also obvious that in flat spacetime owing to a vanishing affine connection the geodesics become straight lines.
\newline
Since $dx^{\mu}$ and $dx^{\nu}$ are symmetric, the contribution to the equation of motion (\ref{45}) comes from just the symmetric part of the affine connection, which is known from Eq.(\ref{33}) to be 
\begin{equation}
\Gamma^{\rho}_{\left(\mu\nu\right)}=\lbrace^{\rho}_{\mu\nu}\rbrace_{\mathcal{R}\left(\Gamma\right)}-\frac{1}{3}\left(\delta_{\mu}^{\rho}\mathcal{S_{\nu}}+\delta_{\nu}^{\rho}\mathcal{S_{\mu}}\right).\label{46}
\end{equation}
Putting Eq.(\ref{46}) into (\ref{45}), we get 
\begin{equation}
\ddot{x}^{\rho}+\lbrace^{\rho}_{\mu\nu}\rbrace_{\mathcal{R}\left(\Gamma\right)}\dot{x}^{\mu}\dot{x}^{\nu}-\frac{2}{3}\mathcal{S_{\mu}}\dot{x}^{\mu}\dot{x}^{\rho}=0,\label{47}
\end{equation}
where $\ddot{x}^{\mu}\equiv \frac{d^{2}x^{\mu}}{ds^{2}}$ and $\dot{x}^{\mu}\equiv \frac{dx^{\mu}}{ds}$. After writing the explicit form of the Christoffel-brackets and using the symmetry of $\dot{x}^{\mu}$ and $\dot{x}^{\nu}$, Eq.(\ref{47}) becomes  
\begin{equation}
\ddot{x}^{\rho}+\left(\mathcal{R}^{-1}\right)^{\rho\sigma}\frac{d \mathcal{R}_{\sigma\mu}}{ds}\dot{x}^{\mu}-\frac{1}{2}\left(\mathcal{R}^{-1}\right)^{\rho\sigma}\left(\partial_{\sigma}\mathcal{R}_{\mu\nu}\right)\dot{x}^{\mu}\dot{x}^{\nu}-\frac{2}{3}\mathcal{S_{\mu}}\dot{x}^{\mu}\dot{x}^{\rho}=0.
\end{equation}
Multiplying the last equation by $\mathcal{R}_{\rho\beta}\dot{x}^{\beta}$ gives
\begin{eqnarray}
\mathcal{R}_{\rho\beta}\dot{x}^{\beta}\frac{d}{ds}\dot{x}^{\rho}+\frac{1}{2}\frac{d\mathcal{R}_{\rho\beta}}{ds}\dot{x}^{\beta}\dot{x}^{\rho}-\frac{2}{3}\mathcal{R}_{\rho\beta}\mathcal{S_{\mu}}\dot{x}^{\mu}\dot{x}^{\beta}\dot{x}^{\rho}\nonumber\\=\frac{d}{ds}\left(\mathcal{R}_{\rho\beta}\dot{x}^{\beta}\dot{x}^{\rho}\right)-\frac{4}{3}\mathcal{R}_{\rho\beta}\mathcal{S_{\mu}}\dot{x}^{\mu}\dot{x}^{\beta}\dot{x}^{\rho}=0.\label{49}
\end{eqnarray}
Finally, dividing by $\mathcal{R}_{\rho\beta}\dot{x}^{\beta}\dot{x}^{\rho}$, Eq.(\ref{49}) leads to
\begin{equation}
d\ln\big(\mathcal{R_{\rho\beta}}\dot{x}^{\beta}\dot{x}^{\rho}\big)=\frac{4}{3}\mathcal{S_{\mu}}dx^{\mu},
\end{equation}
which after integrating results in the relation
\begin{equation}
\mathcal{R}_{\mu\nu}\left(x^{\sigma}\right)dx^{\mu}dx^{\nu}=\exp\left\lbrace\frac{4}{3}\int^{x^{\sigma}}_{x^{\sigma}_{0}}\mathcal{S_{\rho}}dx^{\rho}\right\rbrace\mathcal{R}_{\mu\nu}\left(x^{\sigma}_{0}\right)dx^{\mu}_{0}dx^{\nu}_{0}.
\end{equation}
The importance of the last expression comes from the metrical interpretation of the Ricci tensor proposed by Eddington \cite{E3} such that; with his interpretation, we can identify the line element $d\tilde{s}^{2}=\mathcal{R}_{\mu\nu}\left(x^{\sigma}\right)dx^{\mu}dx^{\nu}$ which is conformally transported in spacetime because of the torsion effect:
\begin{equation}
d\tilde{s}^{2}\left(x^{\sigma}\right)=\exp\left\lbrace\frac{4}{3}\int^{x^{\sigma}}_{x^{\sigma}_{0}}\mathcal{S_{\rho}}dx^{\rho}\right\rbrace d\tilde{s}^{2}\left(x_{0}^{\sigma}\right).
\end{equation} 

After writing the field equations as symmetric and antisymmetric parts, we first considered the case of the symmetric Ricci tensor such that it was the only condition to get and make use of the affine connection (\ref{33}) which led to some results of the affine space of the symetric Ricci tensor. Next, we will see the case where the Ricci tensor is taken antisymmetric, i.e. $\mathcal{R}_{\mu\nu}=\mathcal{R}_{[\mu\nu]}$. In this case, from Eq.(\ref{S2}) we have
\begin{equation}
\nabla_{\rho}\mathcal{R}_{\mu\nu}=\frac{2}{3}\mathcal{S_{\rho}}\mathcal{R}_{\mu\nu}-\frac{1}{3}\mathcal{S_{\mu}}\mathcal{R}_{\nu\rho}+\frac{1}{3}\mathcal{S_{\nu}}\mathcal{R}_{\mu\rho}-\mathcal{S^{\sigma}\newline_{\rho\mu}}\mathcal{R}_{\nu\sigma}+\mathcal{S^{\sigma}\newline_{\rho\nu}}\mathcal{R}_{\mu\sigma}.\label{52}
\end{equation}
Here, by implementing the cyclic permutation, we are allowed to acquire only a contracted connection:
\begin{equation}
\Gamma^{\beta}_{\beta\rho}=\frac{1}{2}\lbrace^{\beta}_{\beta\rho}\rbrace_{\mathcal{R}_{a}\left(\Gamma\right)}-\frac{8}{3}\mathcal{S_{\rho}}.\label{53}
\end{equation}
 We have here denoted another kind of contracted bracket symbol with respect to the antisymmetric Ricci tensor given by
 \begin{equation}
 \lbrace^{\beta}_{\beta\rho}\rbrace_{\mathcal{R}_{a}\left(\Gamma\right)}=\frac{1}{2}\left(\mathcal{R}^{-1}\right)^{\beta\sigma}\left[\partial_{[\beta}\mathcal{R}_{\sigma]\rho}+\partial_{\rho}\mathcal{R}_{\sigma\beta}-\partial_{[\sigma}\mathcal{R}_{\beta]\rho}\right],
 \end{equation}
which has the same form as the contracted Christoffel-brackets with respect to the symmetric Ricci tensor written as
 \begin{equation}
 \lbrace^{\beta}_{\beta\rho}\rbrace_{\mathcal{R}_{s}\left(\Gamma\right)}=\frac{1}{2}\left(\mathcal{R}^{-1}\right)^{\beta\sigma}\left[\partial_{(\beta}\mathcal{R}_{\sigma)\rho}+\partial_{\rho}\mathcal{R}_{\sigma\beta}-\partial_{(\sigma}\mathcal{R}_{\beta)\rho}\right].
 \end{equation} 
 If we look at the rescaling of the antisymmetric Ricci tensor $\mathcal{R}_{\mu\nu}$, we can see from Eq.(\ref{52}) after applying $\left(\mathcal{R}^{-1}\right)^{\nu\mu}$ product [or directly from Eq.(\ref{xz})] that
 \begin{equation}
\mathcal{R}_{\mu\nu}\left(x^{\sigma}\right)=\exp\left\lbrace\int_{x_{0}^{\sigma}}^{x^{\sigma}}\left(\frac{4}{3}\mathcal{S_{\rho}}+\frac{1}{2}\Gamma^{\beta}_{\beta\rho}\right)dx^{\rho}\right\rbrace\mathcal{R}_{\mu\nu}\left(x_{0}^{\sigma}\right),\label{55}
\end{equation}
and plugging (\ref{53}) into the last equation, we again see that the contracted connection eliminates the explicit torsion effects in Eq.(\ref{55}), and then we end up with the non- local, exponential rescaling of the antisymmetric Ricci tensor, where only the contracted Christoffel-brackets are found, as follows:
\begin{equation}
\mathcal{R}_{\mu\nu}\left(x^{\sigma}\right)=\exp\left\lbrace\int_{x_{0}^{\sigma}}^{x^{\sigma}}\frac{1}{4}\lbrace^{\beta}_{\beta\rho}\rbrace_{\mathcal{R}_{a}\left(\Gamma\right)}dx^{\rho}\right\rbrace\mathcal{R}_{\mu\nu}\left(x_{0}^{\sigma}\right),\label{56}
\end{equation}
where the non-local conformal factor is then written as 
\begin{equation}
\mathcal{K}_{a}\left(x^{\sigma},x^{\sigma}_{0}\right)=\exp\left\lbrace\int_{x_{0}^{\sigma}}^{x^{\sigma}}\frac{1}{4}\lbrace^{\beta}_{\beta\rho}\rbrace_{\mathcal{R}\left(\Gamma\right)}dx^{\rho}\right\rbrace.
\end{equation}
 
In addition to all considerations given above, if we apply the same cyclic permutation procedure directly (without any restrictions on the general Ricci tensor) to Eqs.(\ref{S1}) and (\ref{S2}), we can see the relations between the symmetric and antisymmetric parts of the connection  \cite{E1}; Eq.(\ref{S1}) leads to the symmetric part of the connection as a linear function of the antisymmetric or torsional part of the connection:
 \begin{eqnarray}
 \Gamma^{\sigma}_{(\mu\nu)}\mathcal{R}_{(\rho\sigma)}&=\mathcal{S}^{\sigma}\newline_{\mu\rho}\mathcal{R}_{[\sigma\nu]}+\mathcal{S}^{\sigma}\newline_{\nu\rho}\mathcal{R}_{[\sigma\mu]}-\frac{1}{3}\mathcal{S}_{\mu}\mathcal{R}_{\rho\nu}-\frac{1}{3}\mathcal{S}_{\nu}\mathcal{R}_{\rho\mu}\nonumber\\&+\frac{1}{2}\left[\partial_{\mu}\mathcal{R}_{(\nu\rho)}+\partial_{\nu}\mathcal{R}_{(\mu\rho)}-\partial_{\rho}\mathcal{R}_{(\mu\nu)}\right],\label{59}
 \end{eqnarray}
 and Eq.(\ref{S2}) gives the torsion tensor as a linear function of its vector and of the symmetric part of the connection:
 \begin{eqnarray}
 \mathcal{S}^{\sigma}\newline_{\mu\nu}\mathcal{R}_{(\rho\sigma)}&=\Gamma^{\sigma}_{(\mu\rho)}\mathcal{R}_{[\nu\sigma]}+\Gamma^{\sigma}_{(\nu\rho)}\mathcal{R}_{[\sigma\mu]}+\frac{1}{3}\mathcal{S}_{\mu}\mathcal{R}_{\nu\rho}-\frac{1}{3}\mathcal{S}_{\nu}\mathcal{R}_{\mu\rho}-\frac{2}{3}\mathcal{S}_{\rho}\mathcal{R}_{[\mu\nu]}\nonumber\\&+\frac{1}{2}\left[\partial_{\mu}\mathcal{R}_{[\rho\nu]}+\partial_{\nu}\mathcal{R}_{[\mu\rho]}-\partial_{\rho}\mathcal{R}_{[\nu\mu]}\right].\label{60}
 \end{eqnarray}
Equations (\ref{59}) and (\ref{60}) result in the linear combination of the affine connection as a linear function of the torsion vector and of the derivatives of the Ricci tensor: 
\begin{eqnarray}
\Gamma^{\sigma}_{\mu\nu}\mathcal{R}_{(\rho\sigma)}+\Gamma^{\sigma}_{\rho\mu}\mathcal{R}_{[\sigma\nu]}+\Gamma^{\sigma}_{\nu\rho}\mathcal{R}_{[\mu\sigma]}&=\frac{2}{3}\mathcal{S}_{\mu}\mathcal{R}_{[\nu\rho]}-\frac{2}{3}\mathcal{S}_{\nu}\mathcal{R}_{(\mu\rho)}-\frac{2}{3}\mathcal{S}_{\rho}\mathcal{R}_{[\mu\nu]}\nonumber\\&+\frac{1}{2}\left[\partial_{\mu}\mathcal{R}_{\rho\nu}+\partial_{\nu}\mathcal{R}_{\mu\rho}-\partial_{\rho}\mathcal{R}_{\nu\mu}\right].
\end{eqnarray}
As expected, the last equation is also provided directly by the general field equations (\ref{21}) after applying the cyclic permutation.

\section{Torsional extension to Ricci determinant}

In order to see the torsion effects on the affine dynamics more explicitly, we extend our argument with a torsional action, 
\begin{equation}
I_{\mathcal{R}\mathcal{S}}=\int d^{4}x\lbrace\textit{a}\sqrt{\vert\mathcal{R}\vert}+\textit{b}{\vert\mathcal{S}\vert}\rbrace,\label{61}
\end{equation}
where $\mathcal{S}\equiv \texttt{Det}\left[\mathcal{S^{\rho}\newline_{\mu\nu}}\right]$ and the constants $\textit{a}$ and $\textit{b}$ are dimensionless. In the action (\ref{61}), the determinant of the torsion tensor is defined as
\begin{align}
\texttt{Det}[\mathcal{S}^{\rho}\newline_{\mu\nu}]=\Big(&\frac{1}{4!}\epsilon^{\alpha_0\alpha_1\alpha_2\alpha_3}\epsilon^{\beta_0\beta_{1}\beta_{2}\beta_{3}}\nonumber\\&\times\mathcal{S}^{\rho_0}\newline_{\alpha_0\mu_0}\mathcal{S}^{\mu_0}\newline_{\rho_0\beta_0}\mathcal{S}^{\rho_{1}}\newline_{\alpha_{1}\mu_{1}}\mathcal{S}^{\mu_{1}}\newline_{\rho_{1}\beta_{1}}\mathcal{S}^{\rho_{2}}\newline_{\alpha_{2}\mu_{2}}\mathcal{S}^{\mu_{2}}\newline_{\rho_{2}\beta_{2}}\mathcal{S}^{\rho_{3}}\newline_{\alpha_{3}\mu_{3}}\mathcal{S}^{\mu_{3}}\newline_{\rho_{3}\beta_{3}}\Big)^\frac{1}{2},\label{abcdefg}
\end{align}
which is due to the fact that the construction of an even rank tensor as the direct product of an original odd-rank tensor is the recipe to define the determinant of the odd-rank tensor \cite{GMA,V0,V}.
\newline
Varying the action leads to 
\begin{equation}
\delta I_{\mathcal{R}\mathcal{S}}=\frac{\textit{a}}{2}\int d^{4}x\sqrt{\vert\mathcal{R}\vert}\left(\mathcal{R}^{-1}\right)^{\nu\mu}\delta\mathcal{R_{\mu\nu}}+\textit{b}\int d^{4}x\vert\mathcal{S}\vert\left(\mathcal{S}^{-1}\right)^{\nu\mu}\newline_{\rho}\delta\mathcal{S^{\rho}\newline_{\mu\nu}}.\label{62}
\end{equation}
Since we are concerned with the variation with respect to the nonsymmetric gravitational field, we obtain from the second term in Eq.(\ref{62})
\begin{eqnarray}
\textit{b}\int d^{4}x\vert\mathcal{S}\vert\left(\mathcal{S}^{-1}\right)^{\nu\mu}\newline_{\rho}\delta\mathcal{S^{\rho}\newline_{\mu\nu}}&=\frac{\textit{b}}{2}\int d^{4}x\vert\mathcal{S}\vert\left(\mathcal{S}^{-1}\right)^{\nu\mu}\newline_{\rho}\left[\delta\Gamma^{\rho}_{\mu\nu}-\delta\Gamma^{\rho}_{\nu\mu}\right]\nonumber\\&=\frac{\textit{b}}{2}\int d^{4}x\vert\mathcal{S}\vert\left[\left(\mathcal{S}^{-1}\right)^{\nu\mu}\newline_{\rho}-\left(\mathcal{S}^{-1}\right)^{\mu\nu}\newline_{\rho}\right]\delta\Gamma^{\rho}_{\mu\nu}\nonumber\\&=\textit{b}\int d^{4}x\vert\mathcal{S}\vert\left(\mathcal{S}^{-1}\right)^{\nu\mu}\newline_{\rho}\delta\Gamma^{\rho}_{\mu\nu}.\label{63}
\end{eqnarray}
Then adding Eq.(\ref{63}) to the variation (\ref{13}), and keeping our action stationary the most general field equations become
\begin{eqnarray}
\nabla_{\rho}\left[\sqrt{\vert\mathcal{R}\vert}\left(\mathcal{R}^{-1}\right)^{\nu\mu}\right]-\nabla_{\sigma}\left[\sqrt{\vert\mathcal{R}\vert}\left(\mathcal{R}^{-1}\right)^{\sigma\mu}\right]\delta_{\rho}^{\nu}-2\sqrt{\vert\mathcal{R}\vert}\left(\mathcal{R}^{-1}\right)^{\nu\mu}\mathcal{S_{\rho}}\nonumber\\+2\sqrt{\vert\mathcal{R}\vert}\left(\mathcal{R}^{-1}\right)^{\sigma\mu}\mathcal{S_{\sigma}}\delta_{\rho}^{\nu}+2\sqrt{\vert\mathcal{R}\vert}\left(\mathcal{R}^{-1}\right)^{\sigma\mu}\mathcal{S^{\nu}\newline_{\rho\sigma}}-\frac{2\textit{b}}{\textit{a}}\vert\mathcal{S}\vert\left(\mathcal{S}^{-1}\right)^{\nu\mu}\newline_{\rho}\label{64}=0.
\end{eqnarray}
Applying the contraction with respect to the indices $\rho$ and $\nu$ gives the inverse torsion vector $\left(\mathcal{S}^{-1}\right)^{\mu}$  as the contribution to Eq.(\ref{15}):
\begin{equation}
\nabla_{\sigma}\left[\sqrt{\vert\mathcal{R}\vert}\left(\mathcal{R}^{-1}\right)^{\sigma\mu}\right]=\frac{4}{3}\sqrt{\vert\mathcal{R}\vert}\left(\mathcal{R}^{-1}\right)^{\sigma\mu}\mathcal{S_{\sigma}}-\frac{2\textit{b}}{3\textit{a}}\vert\mathcal{S}\vert\left(\mathcal{S}^{-1}\right)^{\mu}.\label{65}
\end{equation}
Then after multiplying Eq.(\ref{64}) by $\mathcal{R}_{\mu\nu}$ and plugging Eq.(\ref{65}) into (\ref{64}) by taking advantage of Eq.(\ref{17}), we obtain
\begin{equation}
\nabla_{\rho}\sqrt{\vert\mathcal{R}\vert}=\frac{8}{3}\sqrt{\vert\mathcal{R}\vert}\mathcal{S_{\rho}}-\frac{b}{a}\vert\mathcal{S}\vert\left(\frac{1}{3}\left(\mathcal{S}^{-1}\right)^{\mu}\delta^{\nu}_{\rho}-\left(\mathcal{S}^{-1}\right)^{\nu\mu}\newline_{\rho}\right)\mathcal{R_{\mu\nu}},\label{66}
\end{equation}
which, by using Eq.(\ref{64}) together with Eq.(\ref{65}), gives
\begin{eqnarray}
\fl\nabla_{\rho}\left[\left(\mathcal{R}^{-1}\right)^{\nu\mu}\right]+\frac{2}{3}\mathcal{S_{\rho}}\left(\mathcal{R}^{-1}\right)^{\nu\mu}+\frac{2}{3}\mathcal{S_{\sigma}}\delta_{\rho}^{\nu}\left(\mathcal{R}^{-1}\right)^{\sigma\mu}+2\mathcal{S^{\nu}\newline_{\rho\sigma}}\left(\mathcal{R}^{-1}\right)^{\sigma\mu}\nonumber\\ \fl-\mathcal{F}\left\lbrace\mathcal{R_{\alpha\beta}}\left[\frac{1}{3}\left(\mathcal{S}^{-1}\right)^{\alpha}\delta^{\beta}_{\rho}-\left(\mathcal{S}^{-1}\right)^{\beta\alpha}\newline_{\rho}\right]\left(\mathcal{R}^{-1}\right)^{\nu\mu}-\frac{2}{3}\left(\mathcal{S}^{-1}\right)^{\mu}\delta^{\nu}_{\rho}+2\left(\mathcal{S}^{-1}\right)^{\nu\mu}\newline_{\rho}\right\rbrace=0,\label{67}
\end{eqnarray}
where $\mathcal{F}\equiv \textit{f}\left(\mathcal{R},\mathcal{S}\right)=\frac{b\vert\mathcal{S}\vert}{a\sqrt{\vert\mathcal{R}\vert}} $ is a scalar function since the torsion and Ricci parts in the action (\ref{61}) are the scalar density of the identical weights $+1$, which means that the ratio of both parts results in a scalar of weight $0$ .
\newline
Lowering the indices of the inverse Ricci tensor in Eq.(\ref{67}) is equivalent to the sufficient final equations for the gravitational field:
\begin{eqnarray}
&\nabla_{\rho}\left[\mathcal{R_{\mu\nu}}\right]-\frac{2}{3}\left[\mathcal{S_{\rho}}-\frac{3\mathcal{F}}{2}\mathcal{R_{\alpha\sigma}}\left(\frac{1}{3}\left(\mathcal{S}^{-1}\right)^{\alpha}\delta^{\sigma}_{\rho}-\left(\mathcal{S}^{-1}\right)^{\sigma\alpha}\newline_{\rho}\right)\right]\mathcal{R_{\mu\nu}}\nonumber\\&-\frac{2}{3}\left[\mathcal{S_{\nu}}+\mathcal{F}\mathcal{R_{\alpha\nu}}\left(\mathcal{S}^{-1}\right)^{\alpha}\right]\mathcal{R_{\mu\rho}}\nonumber\\&-2\left[\mathcal{S^{\sigma}\newline_{\rho\nu}}-\mathcal{F}\mathcal{R_{\alpha\nu}}\left(\mathcal{S}^{-1}\right)^{\sigma\alpha}\newline_{\rho}\right]\mathcal{R_{\mu\sigma}}=0.\label{68}
\end{eqnarray}
If we compare Eq.(\ref{68}) with (\ref{21}) we can see the contributions of the torsion determinant to the case of the purely Ricci determinant such that the contributions come from inverse torsion tensor and vector, which are coupled to the Ricci tensor and scalar function.

Let us now see the effects of the torsional gravitational equations (\ref{68}) on our Ricci tensor such that after $\left(\mathcal{R}^{-1}\right)^{\nu\mu}$ product of Eq.(\ref{68}) [or directly from Eqs.(\ref{66}) and (\ref{17})], we get
\begin{equation}
\left(\mathcal{R}^{-1}\right)^{\nu\mu}\nabla_{\rho}\left[\mathcal{R_{\mu\nu}}\right]=\frac{16}{3}\mathcal{S_{\rho}}-2\mathcal{F}\left(\frac{1}{3}\left(\mathcal{S}^{-1}\right)^{\mu}\delta^{\nu}_{\rho}-\left(\mathcal{S}^{-1}\right)^{\nu\mu}\newline_{\rho}\right)\mathcal{R_{\mu\nu}},
\end{equation}
which leads to
\begin{equation}
\left(\mathcal{R}^{-1}\right)^{\nu\mu}\partial_{\rho}\left[\mathcal{R_{\mu\nu}}\right]=\frac{16}{3}\mathcal{S_{\rho}}+2\Gamma^{\beta}_{\beta\rho}-2\mathcal{F}\left(\frac{1}{3}\left(\mathcal{S}^{-1}\right)^{\mu}\delta^{\nu}_{\rho}-\left(\mathcal{S}^{-1}\right)^{\nu\mu}\newline_{\rho}\right)\mathcal{R_{\mu\nu}}.
\end{equation}
Thus, using the Jacobi formula (\ref{19}), we have
\begin{eqnarray}
\fl\texttt{Det}\left[\mathcal{R_{\mu\nu}}\left(x^{\sigma}\right)\right]=\exp\Bigg\{&\int_{x_{0}^{\sigma}}^{x^{\sigma}}\Bigg[\frac{16}{3}\mathcal{S_{\rho}}+2\Gamma^{\beta}_{\beta\rho}\nonumber\\&- 2\mathcal{F}\Bigg(\frac{1}{3}\left(\mathcal{S}^{-1}\right)^{\alpha}\delta^{\beta}_{\rho}-\left(\mathcal{S}^{-1}\right)^{\beta\alpha}\newline_{\rho}\Bigg)\mathcal{R_{\alpha\beta}}\Bigg]dx^{\rho}\Bigg\}\texttt{Det}\left[\mathcal{R_{\mu\nu}}\left(x_{0}^{\sigma}\right)\right],
\end{eqnarray}
which, via the general transformation of the coordinates, results in the rescaling of Ricci tensor as
\begin{eqnarray}
\fl\mathcal{R_{\mu\nu}}\left(x^{\sigma}\right)=\exp\Bigg\{&\int_{x_{0}^{\sigma}}^{x^{\sigma}}\Bigg[\frac{4}{3}\mathcal{S_{\rho}}+\frac{1}{2}\Gamma^{\beta}_{\beta\rho}\nonumber\\&-\frac{\mathcal{F}}{2}\left(\frac{1}{3}\left(\mathcal{S}^{-1}\right)^{\alpha}\delta^{\beta}_{\rho}-\left(\mathcal{S}^{-1}\right)^{\beta\alpha}\newline_{\rho}\right)\mathcal{R_{\alpha\beta}}\Bigg]dx^{\rho}\Bigg\}\mathcal{R_{\mu\nu}}\left(x_{0}^{\sigma}\right).\label{72}
\end{eqnarray}
It is finally obvious that the non-local rescaling factor is directly given by the exponential term:
\begin{equation}
\mathcal{K}\left(x^{\sigma},x^{\sigma}_{0}\right)=\exp\Bigg\{\int_{x_{0}^{\sigma}}^{x^{\sigma}}\Bigg[\frac{4}{3}\mathcal{S_{\rho}}+\frac{1}{2}\Gamma^{\beta}_{\beta\rho}-\frac{\mathcal{F}}{2}\Bigg(\frac{1}{3}\left(\mathcal{S}^{-1}\right)^{\alpha}\delta^{\beta}_{\rho}-\left(\mathcal{S}^{-1}\right)^{\beta\alpha}\newline_{\rho}\Bigg)\mathcal{R_{\alpha\beta}}\Bigg]dx^{\rho}\Bigg\}.\label{73}
\end{equation}
As clearly understood from the conformal factor (\ref{73}), because of the appearing of Ricci tensor in it, Eq.(\ref{72}) is not an exact solution, that is to say, due to the torsional determinant, the Ricci tensor is also responsible for the conformal transformation of itself. However, after applying some procedures to our field equations (\ref{68}), we will have a simplified solution form, which will be the same as the first action model (\ref{7}).

As we did in the previous section, we can now write the field equations (\ref{68}) as symmetric and antisymmetric parts with respect to the indices $\mu$ and $\nu$:
\begin{eqnarray}
\fl\nabla_{\rho}\mathcal{R_{\left(\mu\nu\right)}}-\frac{2}{3}\left[\mathcal{S_{\rho}}-\frac{3\mathcal{F}}{2}\mathcal{R_{\alpha\sigma}}\left(\frac{1}{3}\left(\mathcal{S}^{-1}\right)^{\alpha}\delta^{\sigma}_{\rho}-\left(\mathcal{S}^{-1}\right)^{\sigma\alpha}\newline_{\rho}\right)\right]\mathcal{R_{\left(\mu\nu\right)}}\nonumber\\\fl-\frac{1}{3}\left[\mathcal{S_{\nu}}+\mathcal{F}\mathcal{R_{\alpha\nu}}\left(\mathcal{S}^{-1}\right)^{\alpha}\right]\mathcal{R_{\mu\rho}}-\frac{1}{3}\left[\mathcal{S_{\mu}}+\mathcal{F}\mathcal{R_{\alpha\mu}}\left(\mathcal{S}^{-1}\right)^{\alpha}\right]\mathcal{R_{\nu\rho}}\nonumber\\\fl-\left[\mathcal{S^{\sigma}\newline_{\rho\nu}}-\mathcal{F}\mathcal{R_{\alpha\nu}}\left(\mathcal{S}^{-1}\right)^{\sigma\alpha}\newline_{\rho}\right]\mathcal{R_{\mu\sigma}}-\left[\mathcal{S^{\sigma}\newline_{\rho\mu}}-\mathcal{F}\mathcal{R_{\alpha\mu}}\left(\mathcal{S}^{-1}\right)^{\sigma\alpha}\newline_{\rho}\right]\mathcal{R_{\nu\sigma}}=0\label{74}
\end{eqnarray}
and 
\begin{eqnarray}
\fl\nabla_{\rho}\mathcal{R_{\left[\mu\nu\right]}}-\frac{2}{3}\left[\mathcal{S_{\rho}}-\frac{3\mathcal{F}}{2}\mathcal{R_{\alpha\sigma}}\left(\frac{1}{3}\left(\mathcal{S}^{-1}\right)^{\alpha}\delta^{\sigma}_{\rho}-\left(\mathcal{S}^{-1}\right)^{\sigma\alpha}\newline_{\rho}\right)\right]\mathcal{R_{\left[\mu\nu\right]}}\nonumber\\\fl-\frac{1}{3}\left[\mathcal{S_{\nu}}+\mathcal{F}\mathcal{R_{\alpha\nu}}\left(\mathcal{S}^{-1}\right)^{\alpha}\right]\mathcal{R_{\mu\rho}}+\frac{1}{3}\left[\mathcal{S_{\mu}}+\mathcal{F}\mathcal{R_{\alpha\mu}}\left(\mathcal{S}^{-1}\right)^{\alpha}\right]\mathcal{R_{\nu\rho}}\nonumber\\\fl-\left[\mathcal{S^{\sigma}\newline_{\rho\nu}}-\mathcal{F}\mathcal{R_{\alpha\nu}}\left(\mathcal{S}^{-1}\right)^{\sigma\alpha}\newline_{\rho}\right]\mathcal{R_{\mu\sigma}}+\left[\mathcal{S^{\sigma}\newline_{\rho\mu}}-\mathcal{F}\mathcal{R_{\alpha\mu}}\left(\mathcal{S}^{-1}\right)^{\sigma\alpha}\newline_{\rho}\right]\mathcal{R_{\nu\sigma}}=0.\label{75}
\end{eqnarray}
Eliminating the antisymmetric part of the Ricci tensor, i.e. $\mathcal{R_{\mu\nu}}=\mathcal{R_{(\mu\nu)}}$, Eq.(\ref{74}) becomes
\begin{eqnarray}
\nabla_{\rho}\mathcal{R}_{\mu\nu}&=\frac{2}{3}\left[\mathcal{S_{\rho}}-\frac{\mathcal{F}}{2}\mathcal{R}_{\alpha\rho}\left(\mathcal{S}^{-1}\right)^{\alpha}\right]\mathcal{R}_{\mu\nu}\nonumber\\&+\frac{1}{3}\left[\mathcal{S_{\nu}}+\mathcal{F}\mathcal{R}_{\alpha\nu}\left(\mathcal{S}^{-1}\right)^{\alpha}\right]\mathcal{R}_{\mu\rho}+\frac{1}{3}\left[\mathcal{S_{\mu}}+\mathcal{F}\mathcal{R}_{\alpha\mu}\left(\mathcal{S}^{-1}\right)^{\alpha}\right]\mathcal{R}_{\nu\rho}\nonumber\\&+\mathcal{S^{\sigma}\newline_{\rho\nu}}\mathcal{R}_{\mu\sigma}+\mathcal{S^{\sigma}\newline_{\rho\mu}}\mathcal{R}_{\nu\sigma}.\label{76}
\end{eqnarray}
From now on, we are ready to obtain a nonsymmetric connection structure by taking advantage of Eq.(\ref{76}) such that 
\begin{eqnarray}
\nabla_{\mu}\mathcal{R}_{\nu\rho}+\nabla_{\nu}\mathcal{R}_{\rho\mu}-\nabla_{\rho}\mathcal{R}_{\mu\nu}&=\mathcal{F}\mathcal{R}_{\alpha\rho}\left(\mathcal{S}^{-1}\right)^{\alpha}\mathcal{R}_{\mu\nu}\nonumber\\&+\frac{2}{3}\left[\mathcal{S_{\mu}}-\frac{\mathcal{F}}{2}\mathcal{R}_{\alpha\mu}\left(\mathcal{S}^{-1}\right)^{\alpha}\right]\mathcal{R}_{\nu\rho}\nonumber\\&+\frac{2}{3}\left[\mathcal{S_{\nu}}-\frac{\mathcal{F}}{2}\mathcal{R}_{\alpha\nu}\left(\mathcal{S}^{-1}\right)^{\alpha}\right]\mathcal{R}_{\mu\rho}\nonumber\\&+2\mathcal{S^{\sigma}\newline_{\mu\rho}}\mathcal{R}_{\nu\sigma}+2\mathcal{S^{\sigma}\newline_{\nu\rho}}\mathcal{R}_{\mu\sigma}.\label{77}
\end{eqnarray}
Thus, Eq.(\ref{77}) with Eq.(\ref{32}) results in the affine connection given by
\begin{eqnarray}
\Gamma^{\rho}_{\mu\nu}&=\lbrace^{\rho}_{\mu\nu}\rbrace_{\mathcal{R}_{s}\left(\Gamma\right)}-\frac{\mathcal{F}}{2}\left(\mathcal{S}^{-1}\right)^{\rho}\mathcal{R}_{\mu\nu}\nonumber\\&-\frac{1}{3}\left[\delta_{\mu}^{\rho}\left(\mathcal{S_{\nu}}-\frac{\mathcal{F}}{2}\mathcal{R}_{\alpha\nu}\left(\mathcal{S}^{-1}\right)^{\alpha}\right)+\delta_{\nu}^{\rho}\left(\mathcal{S_{\mu}}-\frac{\mathcal{F}}{2}\mathcal{R}_{\alpha\mu}\left(\mathcal{S}^{-1}\right)^{\alpha}\right)\right]\nonumber\\&+\mathcal{S^{\rho}\newline_{\mu\nu}},\label{78}
\end{eqnarray}
by which it is clear that due to the additional torsion determinant, the shift from the connection Eq.(\ref{33}) is given by
\begin{equation}
-\frac{\mathcal{F}}{2}\left[\left(\mathcal{S}^{-1}\right)^{\rho}\mathcal{R}_{\mu\nu}-\frac{1}{3}\Bigg(\delta_{\mu}^{\rho}\mathcal{R}_{\alpha\nu}\left(\mathcal{S}^{-1}\right)^{\alpha}+\delta_{\nu}^{\rho}\mathcal{R}_{\alpha\mu}\left(\mathcal{S}^{-1}\right)^{\alpha}\Bigg)\right].
\end{equation}
\newline
Furthermore, multiplying Eq.(\ref{76}) by $\left(\mathcal{R}^{-1}\right)^{\nu\mu}$ or taking the symmetric Ricci tensor from Eq.(\ref{72}) directly, we obtain 
\begin{equation}
\mathcal{R}_{\mu\nu}\left(x^{\sigma}\right)=\exp\left\lbrace\int_{x_{0}^{\sigma}}^{x^{\sigma}}\left[\frac{4}{3}\mathcal{S_{\rho}}+\frac{1}{2}\Gamma^{\beta}_{\beta\rho}-\frac{\mathcal{F}}{6}\left(\mathcal{S}^{-1}\right)^{\alpha}\mathcal{R}_{\alpha\rho}\right]dx^{\rho}\right\rbrace\mathcal{R}_{\mu\nu}\left(x_{0}^{\sigma}\right),\label{80}
\end{equation}
for which we are able to use the connection found in Eq.(\ref{78}) such that
\begin{equation}
\Gamma^{\beta}_{\beta\rho}=\lbrace^{\beta}_{\beta\rho}\rbrace_{\mathcal{R}_{s}\left(\Gamma\right)}-\frac{8}{3}\mathcal{S_{\rho}}+\frac{\mathcal{F}}{3}\left(\mathcal{S}^{-1}\right)^{\alpha}\mathcal{R}_{\alpha\rho}.\label{81}
\end{equation}
By substituting into Eq.(\ref{80}), the contracted connection (\ref{81}) eliminates all explicit torsion contributions in the exponential conformal factor. Thus, we end up with the same mapping form of the symmetric Ricci tensor with our first action model:
\begin{equation}
\mathcal{R}_{\mu\nu}\left(x^{\sigma}\right)=\exp\left\lbrace\int_{x_{0}^{\sigma}}^{x^{\sigma}}\frac{1}{2}\lbrace^{\beta}_{\beta\rho}\rbrace_{\mathcal{R}_{s}\left(\Gamma\right)}dx^{\rho}\right\rbrace\mathcal{R}_{\mu\nu}\left(x_{0}^{\sigma}\right),
\end{equation}
which then has the same non-local rescaling factor form:
\begin{equation}
\mathcal{K}_{s}\left(x^{\sigma},x^{\sigma}_{0}\right)=\exp\left\lbrace\int_{x_{0}^{\sigma}}^{x^{\sigma}}\frac{1}{2}\lbrace^{\beta}_{\beta\rho}\rbrace_{\mathcal{R}_{s}\left(\Gamma\right)}dx^{\rho}\right\rbrace.
\end{equation}
However, with the decomposition we can see that the structure of the symmetric Ricci tensor $\mathcal{R}_{\mu\nu}(\Gamma)$ is now different from the first action model and it is given by
\begin{eqnarray}
\mathcal{R}_{\mu\nu}\left(\Gamma\right)&=\mathcal{R}_{\mu\nu}\left(\lbrace^{\rho}_{\mu\nu}\rbrace_{\mathcal{R}_{s}\left(\Gamma\right)}\right)-\frac{1}{3}\mathcal{S}_{\mu}\mathcal{S}_{\nu}-\mathcal{S}^{\rho}\newline_{\mu\sigma}\mathcal{S}^{\sigma}\newline_{\nu\rho}\nonumber\\&-\frac{1}{2}\nabla_{\rho}\left[\mathcal{F}\left(\mathcal{S}^{-1}\right)^{\rho}\mathcal{R}_{\mu\nu}\right]\nonumber\\&+\frac{\mathcal{F}}{9}\left[\mathcal{S}_{\mu}\left(\mathcal{S}^{-1}\right)^{\sigma}\mathcal{R}_{\sigma\nu}+\mathcal{S}_{\nu}\left(\mathcal{S}^{-1}\right)^{\sigma}\mathcal{R}_{\sigma\mu}\right]\nonumber\\&+\frac{\mathcal{F}^{2}}{18}\left[3\left(\mathcal{S}^{-1}\right)^{\sigma}\left(\mathcal{S}^{-1}\right)^{\rho}\mathcal{R}_{\mu\nu}\mathcal{R}_{\rho\sigma}-5\left(\mathcal{S}^{-1}\right)^{\sigma}\left(\mathcal{S}^{-1}\right)^{\rho}\mathcal{R}_{\rho\mu}\mathcal{R}_{\sigma\nu}\right],
\end{eqnarray}
which then implies that although for the action models (\ref{7}) and (\ref{61}) the forms of the rescalings are the same in the case of the symmetric Ricci tensor, they indeed have different values, that is, the Christoffel-brackets $\lbrace^{\beta}_{\beta\rho}\rbrace_{\mathcal{R}_{s}\left(\Gamma\right)}$ are different in each form due to the differences of the symmetric Ricci tensors.

Next, we examine the geodesic equation such that substituting the symmetric part of affine connection (\ref{78}) into the geodesic equation (\ref{45}) with the same procedure given in the previous section, we get
\begin{eqnarray}
&\ddot{x}^{\rho}+\left(\mathcal{R}^{-1}\right)^{\rho\sigma}\frac{d\mathcal{R}_{\sigma\mu}}{ds}\dot{x}^{\mu}-\frac{1}{2}\left(\mathcal{R}^{-1}\right)^{\rho\sigma}\left(\partial_{\sigma}\mathcal{R}_{\mu\nu}\right)\dot{x}^{\mu}\dot{x}^{\nu}\nonumber\\&-\frac{\mathcal{F}}{2}\left(\mathcal{S}^{-1}\right)^{\rho}\mathcal{R}_{\mu\nu}\dot{x}^{\mu}\dot{x}^{\nu}-\frac{2}{3}\left(\mathcal{S_{\mu}}-\frac{\mathcal{F}}{2}\mathcal{R}_{\alpha\mu}\left(\mathcal{S}^{-1}\right)^{\alpha}\right)\dot{x}^{\mu}\dot{x}^{\rho}=0.\label{85}
\end{eqnarray}
Multiplying by $\mathcal{R}_{\rho\beta}\dot{x}^{\beta}$, Eq.(\ref{85}) becomes
\begin{equation}
\frac{d}{ds}\left(\mathcal{R}_{\rho\beta}\dot{x}^{\beta}\dot{x}^{\rho}\right)-\frac{4}{3}\mathcal{R}_{\rho\beta}\mathcal{S_{\mu}}\dot{x}^{\mu}\dot{x}^{\beta}\dot{x}^{\rho}-\frac{\mathcal{F}}{3}\mathcal{R}_{\rho\beta}\mathcal{R}_{\alpha\mu}\left(\mathcal{S}^{-1}\right)^{\alpha}\dot{x}^{\mu}\dot{x}^{\beta}\dot{x}^{\rho}=0,
\end{equation}
which dividing by $\mathcal{R}_{\rho\beta}\dot{x}^{\beta}\dot{x}^{\rho}$ leads to
\begin{equation}
\mathcal{R}_{\mu\nu}\left(x^{\sigma}\right)dx^{\mu}dx^{\nu}=\exp\Bigg\{\int^{x^{\sigma}}_{x^{\sigma}_{0}}\Bigg[\frac{4}{3}\mathcal{S_{\rho}}+\frac{\mathcal{F}}{3}\mathcal{R}_{\alpha\rho}\left(\mathcal{S}^{-1}\right)^{\alpha}\Bigg]dx^{\rho}\Bigg\}\mathcal{R}_{\mu\nu}\left(x^{\sigma}_{0}\right)dx^{\mu}_{0}dx^{\nu}_{0}.
\end{equation}
If we consider the metrical interpretation of the Ricci tensor, the non-local rescaling of the line element is then written as
\begin{equation}
d\tilde{s}^{2}=\exp\left\lbrace\int^{x^{\sigma}}_{x^{\sigma}_{0}}\left[\frac{4}{3}\mathcal{S_{\rho}}+\frac{\mathcal{F}}{3}\mathcal{R}_{\alpha\rho}\left(\mathcal{S}^{-1}\right)^{\alpha}\right]dx^{\rho}\right\rbrace d\tilde{s}_{0}^{2}.
\end{equation}
Therefore, the torsion vector found in the rescaling factor of the line element in the previous section is here modified by the torsional determinant.

As the final step, let us search the case of the antisymmetric Ricci tensor, i.e. $\mathcal{R_{\mu\nu}}=\mathcal{R_{[\mu\nu]}}$, for which Eq.(\ref{75}) takes the form 
\begin{eqnarray}
\nabla_{\rho}\mathcal{R}_{\mu\nu}&=\frac{2}{3}\left[\mathcal{S_{\rho}}-\frac{3\mathcal{F}}{2}\mathcal{R}_{\alpha\sigma}\left(\frac{1}{3}\left(\mathcal{S}^{-1}\right)^{\alpha}\delta^{\sigma}_{\rho}-\left(\mathcal{S}^{-1}\right)^{\sigma\alpha}\newline_{\rho}\right)\right]\mathcal{R}_{\mu\nu}\nonumber\\&+\frac{1}{3}\left[\mathcal{S_{\nu}}+\mathcal{F}\mathcal{R}_{\alpha\nu}\left(\mathcal{S}^{-1}\right)^{\alpha}\right]\mathcal{R}_{\mu\rho}-\frac{1}{3}\left[\mathcal{S_{\mu}}+\mathcal{F}\mathcal{R}_{\alpha\mu}\left(\mathcal{S}^{-1}\right)^{\alpha}\right]\mathcal{R}_{\nu\rho}\nonumber\\&-2\mathcal{F}\mathcal{R}_{\alpha\nu}\left(\mathcal{S}^{-1}\right)^{\sigma\alpha}\newline_{\rho}\mathcal{R}_{\mu\sigma}+\mathcal{S^{\sigma}\newline_{\rho\nu}}\mathcal{R}_{\mu\sigma}-\mathcal{S^{\sigma}\newline_{\rho\mu}}\mathcal{R}_{\nu\sigma}.\label{89}
\end{eqnarray}
Applying the cyclic permutation, Eq.(\ref{89}) gives the contracted connection:
\begin{equation}
\Gamma^{\beta}_{\beta\rho}=\frac{1}{2}\lbrace^{\beta}_{\beta\rho}\rbrace_{\mathcal{R}_{a}\left(\Gamma\right)}-\frac{8}{3}\mathcal{S_{\rho}}+\mathcal{F}\Bigg(\frac{5}{6}\left(\mathcal{S}^{-1}\right)^{\alpha}\delta^{\beta}_{\rho}-\left(\mathcal{S}^{-1}\right)^{\beta\alpha}\newline_{\rho}\Bigg)\mathcal{R}_{\alpha\beta},\label{90}
\end{equation}
After taking the $\left(\mathcal{R}^{-1}\right)^{\nu\mu}$ product
of Eq.(\ref{89}), we get
\begin{eqnarray}
\fl\mathcal{R}_{\mu\nu}\left(x^{\sigma}\right)=\exp\Bigg\{&\int_{x_{0}^{\sigma}}^{x^{\sigma}}\Bigg[\frac{4}{3}\mathcal{S_{\rho}}+\frac{1}{2}\Gamma^{\beta}_{\beta\rho}\nonumber\\&-\frac{\mathcal{F}}{2}\left(\frac{1}{3}\left(\mathcal{S}^{-1}\right)^{\alpha}\delta^{\beta}_{\rho}-\left(\mathcal{S}^{-1}\right)^{\beta\alpha}\newline_{\rho}\right)\mathcal{R}_{\alpha\beta}\Bigg]dx^{\rho}\Bigg\}\mathcal{R}_{\mu\nu}\left(x_{0}^{\sigma}\right),\label{91}
\end{eqnarray}
which is, as expected, compatible with Eq.(\ref{72}).
\newline
Thus, substituting Eq.(\ref{90}) into (\ref{91}) results in the final expression for the non-local conformal transformation of the antisymmetric Ricci tensor:
\begin{equation}
\mathcal{R}_{\mu\nu}\left(x^{\sigma}\right)=\exp\Bigg\{\int_{x_{0}^{\sigma}}^{x^{\sigma}}\Bigg[\frac{1}{4}\lbrace^{\beta}_{\beta\rho}\rbrace_{\mathcal{R}_{a}\left(\Gamma\right)}+\frac{\mathcal{F}}{4}\left(\mathcal{S}^{-1}\right)^{\beta}\mathcal{R}_{\beta\rho}\Bigg]dx^{\rho}\Bigg\}\mathcal{R}_{\mu\nu}\left(x_{0}^{\sigma}\right),\label{92}
\end{equation}
where the non-local, exponential conformal factor is then written as
\begin{equation}
\mathcal{K}_{a}\left(x^{\sigma},x^{\sigma}_{0}\right)=\exp\Bigg\{\int_{x_{0}^{\sigma}}^{x^{\sigma}}\Bigg[\frac{1}{4}\lbrace^{\beta}_{\beta\rho}\rbrace_{\mathcal{R}_{a}\left(\Gamma\right)}+\frac{\mathcal{F}}{4}\left(\mathcal{S}^{-1}\right)^{\beta}\mathcal{R}_{\beta\rho}\Bigg]dx^{\rho}\Bigg\}.
\end{equation}
In contrast to the case of the symmetric Ricci tensor, the form of the rescaling of the antisymmetric Ricci tensor is now not the same as Eq.(\ref{56}) given for the first action model (\ref{7}), and we can see from Eq.(\ref{92}) that owing to the torsional determinant, the antisymmetric Ricci tensor is also responsible for the conformal mapping of itself from one point to another one.

\section{Riemannian action based on the affine connection}
Up to now we fully examined the Ricci tensor by imposing to it a torsional meaning, and as the Ricci tensor is a subset of Riemann curvatures itself, it is essential to analyze the torsionfull Riemannian action as given below:
\begin{equation}
I_{\mathcal{\Re}}=\int d^{4}x\sqrt{\vert\mathcal{\Re}\vert},\label{94}
\end{equation}
where $\mathcal{\Re}\equiv \texttt{Det}\left[ \mathcal{\Re^{\rho}\newline_{\mu\sigma\nu}}\right]$ and $\mathcal{\Re}\equiv \mathcal{\Re}\left(\Gamma\right)$ with the definition
\begin{eqnarray}
\texttt{Det}[\Re^{\rho}\newline_{\mu\sigma\nu}]&=\frac{1}{(4!)^2}\epsilon_{\rho_{0}\rho_{1}\rho_{2}\rho_{3}}\epsilon^{\mu_0\mu_{1}\mu_{2}\mu_{3}}\epsilon^{\sigma_0\sigma_{1}\sigma_{2}\sigma_{3}}\epsilon^{\nu_0\nu_{1}\nu_{2}\nu_{3}}\nonumber\\&\times\Re^{\rho_0}\newline_{\mu_0\sigma_0\nu_0}\Re^{\rho_{1}}\newline_{\mu_{1}\sigma_{1}\nu_{1}}\Re^{\rho_{2}}\newline_{\mu_{2}\sigma_{2}\nu_{2}}\Re^{\rho_{3}}\newline_{\mu_{3}\sigma_{3}\nu_{3}},\label{102}
\end{eqnarray}
Under variation, we write the action as follows:
\begin{equation}
\delta I_{\mathcal{\Re}}=\frac{1}{2}\int d^{4}x\sqrt{\vert\mathcal{\Re}\vert}\left(\mathcal{\Re}^{-1}\right)^{\nu\sigma\mu}\newline_{\rho}\delta\mathcal{\Re^{\rho}\newline_{\mu\sigma\nu}}
\end{equation}
\newline
In order to examine the dynamics of the action (\ref{94}), we should consider the affine connection field such that using the Palatini formula \cite{E1, P1}, our variation becomes
\begin{equation}
\delta I_{\mathcal{\Re}}=\frac{1}{2}\int d^{4}x\sqrt{\vert\mathcal{\Re}\vert}\left(\mathcal{\Re}^{-1}\right)^{\nu\sigma\mu}\newline_{\rho}\Big[\nabla_{\sigma}\Big( \delta\Gamma^{\rho}_{\mu\nu}\Big)-\nabla_{\nu}\Big( \delta\Gamma^{\rho}_{\mu\sigma}\Big)
-2\mathcal{S^{\lambda}\newline_{\sigma\nu}}\delta\Gamma^{\rho}_{\mu\lambda}\Big].
\end{equation}
After applying integration by parts, by using the identity (\ref{11}) we obtain
\begin{eqnarray}
\fl\delta I_{\mathcal{\Re}}=\int d^{4}x\Bigg(-\frac{1}{2}\nabla_{\sigma}\left[\sqrt{\vert\mathcal{\Re}\vert}\left(\mathcal{\Re}^{-1}\right)^{\nu\sigma\mu}\newline_{\rho}\right]+\frac{1}{2}\nabla_{\sigma}\left[\sqrt{\vert\mathcal{\Re}\vert}\left(\mathcal{\Re}^{-1}\right)^{\sigma\nu\mu}\newline_{\rho}\right]+\sqrt{\vert\mathcal{\Re}\vert}\left(\mathcal{\Re}^{-1}\right)^{\nu\sigma\mu}\newline_{\rho}\mathcal{S_{\sigma}}\nonumber\\-\sqrt{\vert\mathcal{\Re}\vert}\left(\mathcal{\Re}^{-1}\right)^{\sigma\nu\mu}\newline_{\rho}\mathcal{S_{\sigma}}-\sqrt{\vert\mathcal{\Re}\vert}\left(\mathcal{\Re}^{-1}\right)^{\lambda\sigma\mu}\newline_{\rho}\mathcal{S^{\nu}\newline_{\sigma\lambda}}\Bigg)\delta\Gamma^{\rho}_{\mu\nu}.\label{97}
\end{eqnarray}
With the principle $\delta I_{\mathcal{\Re}}=0$, and the fact that the Riemann tensor $\mathcal{\Re^{\rho}\newline_{\mu\sigma\nu}}$ is antisymmetric in its last two indices $\sigma$ and $\nu$, Eq.(\ref{97}) leads to the most general field equations:
\begin{equation}
\nabla_{\sigma}\left[\sqrt{\vert\mathcal{\Re}\vert}\left(\mathcal{\Re}^{-1}\right)^{\nu\sigma\mu}\newline_{\rho}\right]-2\sqrt{\vert\mathcal{\Re}\vert}\left(\mathcal{\Re}^{-1}\right)^{\nu\sigma\mu}\newline_{\rho}\mathcal{S_{\sigma}}+\sqrt{\vert\mathcal{\Re}\vert}\left(\mathcal{\Re}^{-1}\right)^{\lambda\sigma\mu}\newline_{\rho}\mathcal{S^{\nu}\newline_{\sigma\lambda}}=0\label{98}
\end{equation}
We can improve Eq.(\ref{98}) by applying the differentiation by parts for the first term and multiplying it by $\mathcal{\Re^{\rho}\newline_{\mu\alpha\nu}}$ in which processes the Riemann matrix multiplications with its inverse obey the relations $\left(\mathcal{\Re}^{-1}\right)^{\nu\alpha\mu}{\newline_{\rho}}\mathcal{\Re^{\rho}\newline_{\mu\sigma\nu}}=\delta^{\alpha}_{\sigma}$ and $\left(\mathcal{\Re}^{-1}\right)^{\alpha\beta\mu}{\newline_{\rho}}\mathcal{\Re^{\rho}\newline_{\mu\sigma\nu}}=\frac{1}{3}\left(\delta^{\beta}_{\sigma}\delta^{\alpha}_{\nu}-\delta^{\beta}_{\nu}\delta^{\alpha}_{\sigma}\right)$, which is expected owing to the antisymmetric property of the Riemann tensor in its last two indices. Then by the calculations mentioned above we conclude with the relation 
\begin{equation}
\nabla_{\alpha}\left[\sqrt{\vert\mathcal{\Re}\vert}\right]=\frac{4}{3}\sqrt{\vert\mathcal{\Re}\vert}\mathcal{S_{\alpha}}-\sqrt{\vert\mathcal{\Re}\vert}\mathcal{\Re^{\rho}\newline_{\mu\alpha\nu}}\nabla_{\sigma}\left[\left(\mathcal{\Re}^{-1}\right)^{\nu\sigma\mu}\newline_{\rho}\right],
\end{equation}
and using this last expression in the first term of Eq.(\ref{98}), we obtain the field equations without scalar density:
\begin{eqnarray}
\left(\mathcal{\Re}^{-1}\right)^{\nu\sigma\mu}\newline_{\rho}\mathcal{\Re^{\xi}\newline_{\kappa\sigma\beta}}\nabla_{\alpha}\left[\left(\mathcal{\Re}^{-1}\right)^{\beta\alpha\kappa}\newline_{\xi}\right]-\nabla_{\sigma}\left[\left(\mathcal{\Re}^{-1}\right)^{\nu\sigma\mu}\newline_{\rho}\right]\nonumber\\+\frac{2}{3}\left(\mathcal{\Re}^{-1}\right)^{\nu\sigma\mu}\newline_{\rho}\mathcal{S_{\sigma}}-\left(\mathcal{\Re}^{-1}\right)^{\lambda\sigma\mu}\newline_{\rho}\mathcal{S^{\nu}\newline_{\sigma\lambda}}=0.\label{100}
\end{eqnarray}
Due to the Riemann product by its inverse with the common $\sigma$ index in the first term, Eq.(\ref{100}) is not the end for us, that is to say, we can express the Kronecker delta definitions for that term. To do this, it is enough to consider the inverse Riemann expression \cite{GMA, V0, V}
\begin{eqnarray}
\left(\mathcal{\Re}^{-1}\right)^{\nu\sigma\mu}\newline_{\rho}&=\frac{1}{(3!)^2}\frac{1}{\texttt{Det}[\Re^{\rho}\newline_{\mu\sigma\nu}]}\epsilon_{\rho\rho_{1}\rho_{2}\rho_{3}}\epsilon^{\nu\nu_{1}\nu_{2}\nu_{3}}\epsilon^{\sigma\sigma_{1}\sigma_{2}\sigma_{3}}\epsilon^{\mu\mu_{1}\mu_{2}\mu_{3}}\nonumber\\&\times\Re^{\rho_{1}}\newline_{\mu_{1}\sigma_{1}\nu_{1}}\Re^{\rho_{2}}\newline_{\mu_{2}\sigma_{2}\nu_{2}}\Re^{\rho_{3}}\newline_{\mu_{3}\sigma_{3}\nu_{3}}\label{101}
\end{eqnarray}
with the definition of the determinant of the Riemann tensor (\ref{102}). Then multiplying each side of Eq.(\ref{101}) by $\mathcal{\Re^{\xi}\newline_{\kappa\sigma\beta}}$, and considering all possible contractions by taking into account the determinant (\ref{102}), we obtain the relation
\begin{equation}
\left(\mathcal{\Re}^{-1}\right)^{\nu\sigma\mu}\newline_{\rho}\mathcal{\Re^{\xi}\newline_{\kappa\sigma\beta}}=a\delta^{\xi}_{\rho}\delta^{\mu}_{\kappa}\delta^{\nu}_{\beta}+b\delta^{\nu}_{\rho}\delta^{\mu}_{\kappa}\delta^{\xi}_{\beta}+b\delta^{\mu}_{\rho}\delta^{\xi}_{\kappa}\delta^{\nu}_{\beta}+c\delta^{\nu}_{\rho}\delta^{\xi}_{\kappa}\delta^{\mu}_{\beta}+d\delta^{\mu}_{\rho}\delta^{\nu}_{\kappa}\delta^{\xi}_{\beta}+e\delta^{\xi}_{\rho}\delta^{\nu}_{\kappa}\delta^{\mu}_{\beta},\label{103}
\end{equation}
where
$a=\Big(\frac{\Psi+\Phi-2\Theta+12}{360}\Big)$, $b=\Big(\frac{-2\Psi-2\Phi+\Theta+24}{360}\Big)$, $c=\Big(\frac{7\Psi+\Phi-2\Theta-12}{360}\Big)$, $d=\Big(\frac{\Psi+7\Phi-2\Theta-12}{360}\Big)$ and $e=\Big(\frac{-2\Psi-2\Phi+7\Theta}{360}\Big)$
with $\Psi \equiv \left(\mathcal{R}^{-1}\right)^{[\alpha\beta]}\mathcal{Q_{\beta\alpha}}$, $\Phi \equiv \left(\mathcal{Q}^{-1}\right)^{\alpha\beta}\mathcal{R_{[\beta\alpha]}}$, and $\Theta \equiv \left(\mathcal{\Re}^{-1}\right)^{\alpha[\sigma\beta]}{\newline_{\gamma}}\mathcal{\Re^{\gamma}\newline_{[\alpha\sigma]\beta}}$ such that $\mathcal{Q_{\mu\nu}}$ is the rank-two antisymmetric tensor corresponding to the contraction of the Riemann tensor with respect to its first and second indices, $\mathcal{\Re^{\rho}\newline_{\rho\mu\nu}}$, and one may realize that, as we are not interested in a purely symmetric connection, we cannot relate the antisymmetric part of Ricci tensor to $\mathcal{Q_{\mu\nu}}$, i.e. $\mathcal{Q_{\mu\nu}}\neq2\mathcal{R_{[\mu\nu]}}$, for our torsionful gravity.
\newline
One may check Eq.(\ref{103}) by noticing that the first three possible contractions on the right-hand side should, respectively, give $\left(\mathcal{\Re}^{-1}\right)^{\beta\sigma\alpha}{\newline_{\gamma}}\mathcal{\Re^{\gamma}\newline_{\alpha\sigma\beta}}$,  $\left(\mathcal{R}^{-1}\right)^{\alpha\beta}\mathcal{R_{\beta\alpha}}$, and $\left(\mathcal{Q}^{-1}\right)^{\alpha\beta}\mathcal{Q_{\beta\alpha}}$, and these multiplications are naturally equal to 4 which can also be seen directly from Eqs. (\ref{101}) and (\ref{102}); however, for the last three possibilities the situation is now different, so that the last three possible contractions should, respectively, lead to $\Psi$, $\Phi$, and $\Theta$, which means that our results for now are not scalars like 4 but locally scalar functions. Expecting all these results of the contractions, one can also obtain the expressions for $a$, $b$, $c$, $d$, and $e$.
\newline
After the explanations given above, we are now ready to substitute Eq.(\ref{103}) into Eq.(\ref{100}), and hence, our field equations get the more useful form
\begin{eqnarray}
\fl(a-1)\nabla_{\sigma}\Big[\left(\mathcal{\Re}^{-1}\right)^{\nu\sigma\mu}\newline_{\rho}\Big]+e\nabla_{\sigma}\Big[\left(\mathcal{\Re}^{-1}\right)^{\mu\sigma\nu}\newline_{\rho}\Big]-\delta^{\nu}_{\rho}\Big\{b\nabla_{\sigma}\Big[\left(\mathcal{R}^{-1}\right)^{\sigma\mu}\Big]+c\nabla_{\sigma}\Big[\left(\mathcal{Q}^{-1}\right)^{\sigma\mu}\Big]\Big\}\nonumber\\ \fl-\delta^{\mu}_{\rho}\Big\{d\nabla_{\sigma}\Big[\left(\mathcal{R}^{-1}\right)^{\sigma\nu}\Big]+b\nabla_{\sigma}\Big[\left(\mathcal{Q}^{-1}\right)^{\sigma\nu}\Big]\Big\}+\frac{2}{3}\left(\mathcal{\Re}^{-1}\right)^{\nu\sigma\mu}\newline_{\rho}\mathcal{S_{\sigma}}-\left(\mathcal{\Re}^{-1}\right)^{\lambda\sigma\mu}\newline_{\rho}\mathcal{S^{\nu}\newline_{\sigma\lambda}}=0.\label{104}
\end{eqnarray}
After applying the contractions with respect to indices $\rho$, $\nu$ first and $\rho$, $\mu$ second to the last equation, we, respectively, obtain
\begin{eqnarray}
\fl\tilde{a}\nabla_{\sigma}\Big[\left(\mathcal{R}^{-1}\right)^{\sigma\mu}\Big]+\tilde{b}\nabla_{\sigma}\Big[\left(\mathcal{Q}^{-1}\right)^{\sigma\mu}\Big]-\frac{2}{3}\left(\mathcal{R}^{-1}\right)^{\sigma\mu}\mathcal{S_{\sigma}}-\left(\mathcal{\Re}^{-1}\right)^{\lambda\sigma\mu}\newline_{\rho}\mathcal{S^{\rho}\newline_{\sigma\lambda}}=0\label{105}
\end{eqnarray}
\begin{eqnarray}
\fl\tilde{c}\nabla_{\sigma}\Big[\left(\mathcal{R}^{-1}\right)^{\sigma\mu}\Big]+\tilde{d}\nabla_{\sigma}\Big[\left(\mathcal{Q}^{-1}\right)^{\sigma\mu}\Big]-\frac{2}{3}\left(\mathcal{Q}^{-1}\right)^{\sigma\mu}\mathcal{S_{\sigma}}-\left(\mathcal{Q}^{-1}\right)^{\lambda\sigma}\mathcal{S^{\mu}\newline_{\sigma\lambda}}=0\label{106}
\end{eqnarray}
with $\tilde{a}=\Big(\frac{\Psi+44}{60}\Big)$, $\tilde{b}=\Big(\frac{-\Psi+1}{15}\Big)$, $\tilde{c}=\Big(\frac{-\Phi+1}{15}\Big)$, and $\tilde{d}=\Big(\frac{\Phi+44}{60}\Big)$.
\newline
Equations (\ref{105}) and (\ref{106}) can be solved to obtain two field equations as follows:
\begin{eqnarray}
\nabla_{\sigma}\Big[\left(\mathcal{R}^{-1}\right)^{\sigma\mu}\Big]&-\frac{2}{3}\tilde{\tilde{d}}\left(\mathcal{R}^{-1}\right)^{\sigma\mu}\mathcal{S_{\sigma}}-\tilde{\tilde{d}}\left(\mathcal{\Re}^{-1}\right)^{\lambda\sigma\mu}\newline_{\rho}\mathcal{S^{\rho}\newline_{\sigma\lambda}}\nonumber\\&+\frac{2}{3}\tilde{\tilde{b}}\left(\mathcal{Q}^{-1}\right)^{\sigma\mu}\mathcal{S_{\sigma}}+\tilde{\tilde{b}}\left(\mathcal{Q}^{-1}\right)^{\lambda\sigma}\mathcal{S^{\mu}\newline_{\sigma\lambda}}=0
\end{eqnarray}\label{0}
and
\begin{eqnarray}
\nabla_{\sigma}\Big[\left(\mathcal{Q}^{-1}\right)^{\sigma\mu}\Big]&-\frac{2}{3}\tilde{\tilde{a}}\left(\mathcal{Q}^{-1}\right)^{\sigma\mu}\mathcal{S_{\sigma}}-\tilde{\tilde{a}}\left(\mathcal{Q}^{-1}\right)^{\lambda\sigma}\mathcal{S^{\mu}\newline_{\sigma\lambda}}\nonumber\\&+\frac{2}{3}\tilde{\tilde{c}}\left(\mathcal{R}^{-1}\right)^{\sigma\mu}\mathcal{S_{\sigma}}+\tilde{\tilde{c}}\left(\mathcal{\Re}^{-1}\right)^{\lambda\sigma\mu}\newline_{\rho}\mathcal{S^{\rho}\newline_{\sigma\lambda}}=0\label{108}
\end{eqnarray}
in which we have the equations $\tilde{\tilde{a}} \equiv \frac{\tilde{a}}{\tilde{a}\tilde{d}-\tilde{b}\tilde{c}}$, $\tilde{\tilde{b}} \equiv \frac{\tilde{b}}{\tilde{a}\tilde{d}-\tilde{b}\tilde{c}}$, $\tilde{\tilde{c}} \equiv \frac{\tilde{c}}{\tilde{a}\tilde{d}-\tilde{b}\tilde{c}}$, and $\tilde{\tilde{d}} \equiv \frac{\tilde{d}}{\tilde{a}\tilde{d}-\tilde{b}\tilde{c}}$. Thus, by putting the last two equations into Eq.(\ref{104}) we get the final expression for the field equations as given below:

 \begin{eqnarray}
\fl(a-1)\nabla_{\sigma}\Big[\left(\mathcal{\Re}^{-1}\right)^{\nu\sigma\mu}\newline_{\rho}\Big]+e\nabla_{\sigma}\Big[\left(\mathcal{\Re}^{-1}\right)^{\mu\sigma\nu}\newline_{\rho}\Big]-\delta^{\nu}_{\rho}
\Big(\frac{2}{3}\overline{a}\left(\mathcal{R}^{-1}\right)^{\sigma\mu}\mathcal{S_{\sigma}}+\frac{2}{3}\overline{b}\left(\mathcal{Q}^{-1}\right)^{\sigma\mu}\mathcal{S_{\sigma}}\nonumber\\ \fl+\overline{b}\left(\mathcal{Q}^{-1}\right)^{\lambda\sigma}\mathcal{S^{\mu}\newline_{\sigma\lambda}}+\overline{a}\left(\mathcal{\Re}^{-1}\right)^{\lambda\sigma\mu}\newline_{\rho}\mathcal{S^{\rho}\newline_{\sigma\lambda}}\Big)-\delta^{\mu}_{\rho}
\Big(\frac{2}{3}\overline{c}\left(\mathcal{R}^{-1}\right)^{\sigma\nu}\mathcal{S_{\sigma}}+\frac{2}{3}\overline{d}\left(\mathcal{Q}^{-1}\right)^{\sigma\nu}\mathcal{S_{\sigma}}\nonumber\\ \fl+\overline{d}\left(\mathcal{Q}^{-1}\right)^{\lambda\sigma}\mathcal{S^{\nu}\newline_{\sigma\lambda}}+\overline{c}\left(\mathcal{\Re}^{-1}\right)^{\lambda\sigma\nu}\newline_{\rho}\mathcal{S^{\rho}\newline_{\sigma\lambda}}\Big)+\frac{2}{3}\left(\mathcal{\Re}^{-1}\right)^{\nu\sigma\mu}\newline_{\rho}\mathcal{S_{\sigma}}-\left(\mathcal{\Re}^{-1}\right)^{\lambda\sigma\mu}\newline_{\rho}\mathcal{S^{\nu}\newline_{\sigma\lambda}}=0,\label{109}
\end{eqnarray}
where $\overline{a} \equiv \frac{b\tilde{d}-c\tilde{c}}{\tilde{a}\tilde{d}-\tilde{b}\tilde{c}}$, $\overline{b} \equiv \frac{c\tilde{a}-b\tilde{b}}{\tilde{a}\tilde{d}-\tilde{b}\tilde{c}}$, $\overline{c} \equiv \frac{d\tilde{d}-b\tilde{c}}{\tilde{a}\tilde{d}-\tilde{b}\tilde{c}}$, and $\overline{d} \equiv \frac{b\tilde{a}-d\tilde{b}}{\tilde{a}\tilde{d}-\tilde{b}\tilde{c}}$.
\newline 
Here, we can continue considering our most general field equations such that from Eq.(\ref{98}) we are able to state that
\begin{equation}
\nabla_{\sigma}\left[\left(\mathcal{\Re}^{-1}\right)^{\nu\sigma\mu}\newline_{\rho}\right]=-\frac{\nabla_{\sigma}\left[\sqrt{\vert\mathcal{\Re}\vert}\right]}{\left[\sqrt{\vert\mathcal{\Re}\vert}\right]}\left(\mathcal{\Re}^{-1}\right)^{\nu\sigma\mu}\newline_{\rho}+2\left(\mathcal{\Re}^{-1}\right)^{\nu\sigma\mu}\newline_{\rho}\mathcal{S_{\sigma}}-\left(\mathcal{\Re}^{-1}\right)^{\lambda\sigma\mu}\newline_{\rho}\mathcal{S^{\nu}\newline_{\sigma\lambda}}\label{110}
\end{equation}
and by changing the order of the indices $\nu$ and $\mu$ we also have
\begin{equation}
\nabla_{\sigma}\left[\left(\mathcal{\Re}^{-1}\right)^{\mu\sigma\nu}\newline_{\rho}\right]=-\frac{\nabla_{\sigma}\left[\sqrt{\vert\mathcal{\Re}\vert}\right]}{\left[\sqrt{\vert\mathcal{\Re}\vert}\right]}\left(\mathcal{\Re}^{-1}\right)^{\mu\sigma\nu}\newline_{\rho}+2\left(\mathcal{\Re}^{-1}\right)^{\mu\sigma\nu}\newline_{\rho}\mathcal{S_{\sigma}}-\left(\mathcal{\Re}^{-1}\right)^{\lambda\sigma\nu}\newline_{\rho}\mathcal{S^{\mu}\newline_{\sigma\lambda}}.\label{111}
\end{equation}
Thus, by plugging Eqs.(\ref{110}) and (\ref{111}) into Eq.(\ref{109}) and then by multiplying the result by $\mathcal{\Re^{\rho}\newline_{\mu\alpha\nu}}$, where we use the fact that $\left(\mathcal{\Re}^{-1}\right)^{\mu\sigma\nu}{\newline_{\rho}}\mathcal{\Re^{\rho}\newline_{\mu\alpha\nu}}=\frac{\Theta}{4}\delta^{\sigma}_{\alpha}$ 
and $\left(\mathcal{\Re}^{-1}\right)^{\lambda\sigma\nu}{\newline_{\rho}}\mathcal{\Re^{\rho}\newline_{\mu\alpha\nu}}=\frac{\Theta}{12}\left(\delta^{\lambda}_{\mu}\delta^{\sigma}_{\alpha}-\delta^{\lambda}_{\alpha}\delta^{\sigma}_{\mu}\right)$ with the same notion given in Eq.(\ref{103}), we obtain the equation
\begin{eqnarray}
\frac{\nabla_{\alpha}\left[\sqrt{\vert\mathcal{\Re}\vert}\right]}{\left[\sqrt{\vert\mathcal{\Re}\vert}\right]}&=2A\mathcal{S_{\alpha}}- 2\Big(\overline{\overline{a}}\mathcal{R_{\mu\alpha}}\left(\mathcal{\Re}^{-1}\right)^{\lambda\sigma\mu}\newline_{\rho}+2\overline{\overline{b}}\mathcal{R_{\rho\alpha}}\left(\mathcal{Q}^{-1}\right)^{\lambda\sigma}\nonumber\\& +2\overline{\overline{c}}\mathcal{Q_{\mu\alpha}}\left(\mathcal{\Re}^{-1}\right)^{\lambda\sigma\mu}\newline_{\rho}+\overline{\overline{d}}\mathcal{Q_{\rho\alpha}}\left(\mathcal{Q}^{-1}\right)^{\lambda\sigma}\Big)\mathcal{S^{\rho}\newline_{\sigma\lambda}}\label{yz}
\end{eqnarray}
with $\overline{\overline{a}} \equiv 2\frac{b\tilde{d}-c\tilde{c}}{(\tilde{a}\tilde{d}-\tilde{b}\tilde{c})(4-4a-e\Theta)}$, $\overline{\overline{b}} \equiv \frac{c\tilde{a}-b\tilde{b}}{(\tilde{a}\tilde{d}-\tilde{b}\tilde{c})(4-4a-e\Theta)}$, $\overline{\overline{c}} \equiv \frac{d\tilde{d}-b\tilde{c}}{(\tilde{a}\tilde{d}-\tilde{b}\tilde{c})(4-4a-e\Theta)}$, $\overline{\overline{d}} \equiv 2\frac{b\tilde{a}-d\tilde{b}}{(\tilde{a}\tilde{d}-\tilde{b}\tilde{c})(4-4a-e\Theta)}$, and $A=\frac{2}{3}\big(1-\frac{2(b\tilde{d}-c\tilde{c}+b\tilde{a}-d\tilde{b})+\frac{1}{2}(c\tilde{a}-b\tilde{b}+d\tilde{d}-b\tilde{c})}{(\tilde{a}\tilde{d}-\tilde{b}\tilde{c})(4-4a-e\Theta)}\big)$.
\newline
Furthermore, by using Eq.(\ref{12}) and applying the Jacobi formula (\ref{19}), Eq.(\ref{yz}) leads to the determinant relation for the Riemann curvatures at different points:
\begin{eqnarray}
\fl\texttt{Det}[\Re^{\rho}\newline_{\mu\sigma\nu}(x^{\sigma})]=\exp\Bigg\{\int_{x_{0}^{\sigma}}^{x^{\sigma}}\Big[4A\mathcal{S_{\alpha}}-4\Big(\overline{\overline{a}}\mathcal{R_{\mu\alpha}}\left(\mathcal{\Re}^{-1}\right)^{\lambda\sigma\mu}\newline_{\rho}+2\overline{\overline{b}}\mathcal{R_{\rho\alpha}}\left(\mathcal{Q}^{-1}\right)^{\lambda\sigma}\nonumber\\\fl+2\overline{\overline{c}}\mathcal{Q_{\mu\alpha}}\left(\mathcal{\Re}^{-1}\right)^{\lambda\sigma\mu}\newline_{\rho}+\overline{\overline{d}}\mathcal{Q_{\rho\alpha}}\left(\mathcal{Q}^{-1}\right)^{\lambda\sigma}\Big)\mathcal{S^{\rho}\newline_{\sigma\lambda}}+2\Gamma^{\beta}_{\beta\alpha}\Big]dx^{\alpha}\Bigg\}\texttt{Det}[\Re^{\rho}\newline_{\mu\sigma\nu}(x^{\sigma}_{0})].\label{ax}
\end{eqnarray}
We are now in a position to consider that under the general coordinate transformations  $x^{\sigma}_{0}\rightarrow x^{\sigma}(x^{\sigma}_{0})$ the Riemann curvature transforms as

\begin{equation}
\Re^{\rho}\newline_{\mu\sigma\nu}(x^{\sigma})=
\frac{\partial x_{{0}_{\alpha}}}{\partial x_{\rho}}
\frac{\partial x^{\beta}_{0}}{\partial x^{\mu}}\frac{\partial x^{\kappa}_{0}}{\partial x^{\sigma}}\frac{\partial x^{\gamma}_{0}}{\partial x^{\nu}}\Re^{\alpha}\newline_{\beta\kappa\gamma}(x^{\sigma}_{0}),\label{ab}
\end{equation}
which, after taking the determinant of each side and comparing with Eq.(\ref{ax}), gives the simple relation for the transformation of coordinates as given below: 
\begin{eqnarray}
\mathrm{J}^{\mu}_{\nu}=\delta^{\mu}_{\nu}&\exp\Bigg\{\int_{x_{0}^{\sigma}}^{x^{\sigma}}\Big[\frac{A}{4}\mathcal{S_{\alpha}}-\frac{1}{4}\Big(\overline{\overline{a}}\mathcal{R_{\mu\alpha}}\left(\mathcal{\Re}^{-1}\right)^{\lambda\sigma\mu}\newline_{\rho}+2\overline{\overline{b}}\mathcal{R_{\rho\alpha}}\left(\mathcal{Q}^{-1}\right)^{\lambda\sigma}\nonumber\\&+2\overline{\overline{c}}\mathcal{Q_{\mu\alpha}}\left(\mathcal{\Re}^{-1}\right)^{\lambda\sigma\mu}\newline_{\rho}+\overline{\overline{d}}\mathcal{Q_{\rho\alpha}}\left(\mathcal{Q}^{-1}\right)^{\lambda\sigma}\Big)\mathcal{S^{\rho}\newline_{\sigma\lambda}}+\frac{1}{8}\Gamma^{\beta}_{\beta\alpha}\Big]dx^{\alpha}\Bigg\}.\label{cz}
\end{eqnarray}
As the final step, using Eq.(\ref{cz}) in Eq.(\ref{ab}) results in the final expression for the ￼￼Riemannian curvature as the non-local, exponential rescaling of it:

\begin{eqnarray}
\fl\mathcal{\Re^{\rho}\newline_{\mu\sigma\nu}}\left(x^{\sigma}\right)=\exp\Bigg\{\int_{x_{0}^{\sigma}}^{x^{\sigma}}\Big[A\mathcal{S_{\alpha}}-\Big(\overline{\overline{a}}\mathcal{R_{\mu\alpha}}\left(\mathcal{\Re}^{-1}\right)^{\lambda\sigma\mu}\newline_{\rho}+2\overline{\overline{b}}\mathcal{R_{\rho\alpha}}\left(\mathcal{Q}^{-1}\right)^{\lambda\sigma}\nonumber\\+2\overline{\overline{c}}\mathcal{Q_{\mu\alpha}}\left(\mathcal{\Re}^{-1}\right)^{\lambda\sigma\mu}\newline_{\rho}+\overline{\overline{d}}\mathcal{Q_{\rho\alpha}}\left(\mathcal{Q}^{-1}\right)^{\lambda\sigma}\Big)\mathcal{S^{\rho}\newline_{\sigma\lambda}}+\frac{1}{2}\Gamma^{\beta}_{\beta\alpha}\Big]dx^{\alpha}\Bigg\}\mathcal{\Re^{\rho}\newline_{\mu\sigma\nu}}\left(x_{0}^{\sigma}\right).\label{114}
\end{eqnarray}
Thus, we can conclude that all fundamental tensors given in the gravity, $\mathcal{R_{\mu\nu}}$, $\mathcal{Q_{\mu\nu}}$, $\mathcal{S_{\mu}}$, $\mathcal{R^{\rho}\newline_{\mu\sigma\nu}}$, and $\mathcal{S^{\rho}\newline_{\mu\nu}}$, where the first two of them are the subsets of the Riemann curvature and the third one is of the torsion tensor, are now responsible for the non-local conformal transformations of the Riemann curvature such that in the exponential rescaling factor they are coupled to each other with the inverses of some of them, and they are also found as a coupling to construct  locally scalar functions included in $A$, $\overline{\overline{a}}$, $\overline{\overline{b}}$, $\overline{\overline{c}}$, and $\overline{\overline{d}}$. One may also take care that the mapping of Riemann tensor differs from the Ricci one (\ref{xz}) with the appearance of the torsion tensor in the conformal factor of (\ref{114}).

\section{Riemannian action with torsion determinant}

In this last section, since the Riemann and torsion are the most fundamental two tensors in gravity we will give some results of an action involving a torsion determinant as a contribution to the curvature as follows:
\begin{equation}
I_{\mathcal{\Re}\mathcal{S}}=\int d^{4}x\lbrace\textit{a}^\prime\sqrt{\vert\mathcal{\Re}\vert}+\textit{b}{\vert\mathcal{S}\vert}\rbrace,\label{115}
\end{equation}
where $\textit{a}^\prime$ and $\textit{b}$ are dimensionless constants. Applying the variation gives
\begin{equation}
\delta I_{\mathcal{\Re}\mathcal{S}}=\frac{\textit{a}\prime}{2}\int d^{4}x\sqrt{\vert\mathcal{\Re}\vert}\left(\mathcal{\Re}^{-1}\right)^{\nu\sigma\mu}\newline_{\rho}\delta\mathcal{\Re^{\rho}\newline_{\mu\sigma\nu}}+\textit{b}\int d^{4}x\vert\mathcal{S}\vert\left(\mathcal{S}^{-1}\right)^{\nu\mu}\newline_{\rho}\delta\mathcal{S^{\rho}\newline_{\mu\nu}},
\end{equation}
by which we are then able to write the most general field equations as the modification of the torsional determinant to Eq.(\ref{98}): 
\begin{eqnarray}
\nabla_{\sigma}\left[\sqrt{\vert\mathcal{\Re}\vert}\left(\mathcal{\Re}^{-1}\right)^{\nu\sigma\mu}\newline_{\rho}\right]-2\sqrt{\vert\mathcal{\Re}\vert}\left(\mathcal{\Re}^{-1}\right)^{\nu\sigma\mu}\newline_{\rho}\mathcal{S_{\sigma}}\nonumber\\+\sqrt{\vert\mathcal{\Re}\vert}\left(\mathcal{\Re}^{-1}\right)^{\lambda\sigma\mu}\newline_{\rho}\mathcal{S^{\nu}\newline_{\sigma\lambda}}-\frac{b}{a\prime}\vert\mathcal{S}\vert\left(\mathcal{S}^{-1}\right)^{\nu\mu}\newline_{\rho}=0.\label{117}
\end{eqnarray}
Here, following the same processes given in the previous section we obtain
\begin{equation}
\nabla_{\alpha}\left[\sqrt{\vert\mathcal{\Re}\vert}\right]=\frac{4}{3}\sqrt{\vert\mathcal{\Re}\vert}\mathcal{S_{\alpha}}+\frac{b}{a\prime}\vert\mathcal{S}\vert\mathcal{\Re^{\rho}\newline_{\mu\alpha\nu}}\left(\mathcal{S}^{-1}\right)^{\nu\mu}\newline_{\rho}-\sqrt{\vert\mathcal{\Re}\vert}\mathcal{\Re^{\rho}\newline_{\mu\alpha\nu}}\nabla_{\sigma}\left[\left(\mathcal{\Re}^{-1}\right)^{\nu\sigma\mu}\newline_{\rho}\right].\label{118}
\end{equation}
Then by substituting Eq.(\ref{118}) into Eq.(\ref{117}), the field equations take the form as
\begin{eqnarray}
\fl\left(\mathcal{\Re}^{-1}\right)^{\nu\sigma\mu}\newline_{\rho}\mathcal{\Re^{\xi}\newline_{\kappa\sigma\beta}}\nabla_{\alpha}\left[\left(\mathcal{\Re}^{-1}\right)^{\beta\alpha\kappa}\newline_{\xi}\right]-\nabla_{\sigma}\left[\left(\mathcal{\Re}^{-1}\right)^{\nu\sigma\mu}\newline_{\rho}\right]+\frac{2}{3}\left(\mathcal{\Re}^{-1}\right)^{\nu\sigma\mu}\newline_{\rho}\mathcal{S_{\sigma}}\nonumber\\ \fl -\left(\mathcal{\Re}^{-1}\right)^{\lambda\sigma\mu}\newline_{\rho}\mathcal{S^{\nu}\newline_{\sigma\lambda}}-\mathcal{F^{\prime}}\Bigg(\left(\mathcal{\Re}^{-1}\right)^{\nu\sigma\mu}\newline_{\rho}\mathcal{\Re^{\xi}\newline_{\kappa\sigma\beta}}\left(\mathcal{S}^{-1}\right)^{\beta\kappa}\newline_{\xi}-\left(\mathcal{S}^{-1}\right)^{\nu\mu}\newline_{\rho}\Bigg)=0 \label{119}
\end{eqnarray}
with $\mathcal{F^{\prime}}\equiv \textit{f}^{\prime}\left(\mathcal{\Re},\mathcal{S}\right)=\frac{b\vert\mathcal{S}\vert}{a^{\prime}\sqrt{\vert\mathcal{\Re}\vert}}$ being a scalar function. Equation (\ref{103}) is now necessary to improve our last equation such that after using (\ref{103}) in Eq.(\ref{119}), we end up with the modified equation of (\ref{104}) due to the torsional determinant:
\begin{eqnarray}
\fl(a-1)\nabla_{\sigma}\Big[\left(\mathcal{\Re}^{-1}\right)^{\nu\sigma\mu}\newline_{\rho}\Big]+e\nabla_{\sigma}\Big[\left(\mathcal{\Re}^{-1}\right)^{\mu\sigma\nu}\newline_{\rho}\Big]-\delta^{\nu}_{\rho}\Big\{b\nabla_{\sigma}\Big[\left(\mathcal{R}^{-1}\right)^{\sigma\mu}\Big]+c\nabla_{\sigma}\Big[\left(\mathcal{Q}^{-1}\right)^{\sigma\mu}\Big]\Big\}\nonumber\\ \fl-\delta^{\mu}_{\rho}\Big\{d\nabla_{\sigma}\Big[\left(\mathcal{R}^{-1}\right)^{\sigma\nu}\Big]+b\nabla_{\sigma}\Big[\left(\mathcal{Q}^{-1}\right)^{\sigma\nu}\Big]\Big\}+\frac{2}{3}\left(\mathcal{\Re}^{-1}\right)^{\nu\sigma\mu}\newline_{\rho}\mathcal{S_{\sigma}}-\left(\mathcal{\Re}^{-1}\right)^{\lambda\sigma\mu}\newline_{\rho}\mathcal{S^{\nu}\newline_{\sigma\lambda}}\nonumber\\ \fl -\mathcal{F^{\prime}}\Big[B\left(\mathcal{S}^{-1}\right)^{\nu\mu}\newline_{\rho}+C\delta^{\nu}_{\rho}\left(\mathcal{S}^{-1}\right)^{\mu}+D\delta^{\mu}_{\rho}\left(\mathcal{S}^{-1}\right)^{\nu}\Big]=0,\label{120}
\end{eqnarray}
where $B=\Big(\frac{\Psi+\Phi-3\Theta-116}{120}\Big)$, $C=\Big(\frac{-3\Psi-\Phi+\Theta+12}{120}\Big) $, and $D=\Big(\frac{\Psi+3\Phi-\Theta-12}{120}\Big)$. Then after applying two contractions with respect to the indices $\rho$, $\nu$ first and $\rho$, $\mu$ second for the last equation, we obtain two expressions in which Eqs.(\ref{105}) and (\ref{106}) is modified by, respectively, $-\tilde{C}\mathcal{F^{\prime}}\left(\mathcal{S}^{-1}\right)^{\mu}$ and $-\tilde{D}\mathcal{F^{\prime}}\left(\mathcal{S}^{-1}\right)^{\mu}$ with $\tilde{C}=\Big(\frac{-\Psi-8}{12}\Big)$ and $\tilde{D}=\Big(\frac{\Phi+8}{12}\Big)$ such that from these modified equations we are able to find 
\begin{eqnarray}
\fl\nabla_{\sigma}\Big[\left(\mathcal{R}^{-1}\right)^{\sigma\mu}\Big]-\frac{2}{3}\tilde{\tilde{d}}\left(\mathcal{R}^{-1}\right)^{\sigma\mu}\mathcal{S_{\sigma}}-\tilde{\tilde{d}}\left(\mathcal{\Re}^{-1}\right)^{\lambda\sigma\mu}\newline_{\rho}\mathcal{S^{\rho}\newline_{\sigma\lambda}}\nonumber\\+\frac{2}{3}\tilde{\tilde{b}}\left(\mathcal{Q}^{-1}\right)^{\sigma\mu}\mathcal{S_{\sigma}}+\tilde{\tilde{b}}\left(\mathcal{Q}^{-1}\right)^{\lambda\sigma}\mathcal{S^{\mu}\newline_{\sigma\lambda}}+\tilde{\tilde{D}}\mathcal{F^{\prime}}\left(\mathcal{S}^{-1}\right)^{\mu}=0\label{121}
\end{eqnarray}
and
\begin{eqnarray}
\fl\nabla_{\sigma}\Big[\left(\mathcal{Q}^{-1}\right)^{\sigma\mu}\Big]-\frac{2}{3}\tilde{\tilde{a}}\left(\mathcal{Q}^{-1}\right)^{\sigma\mu}\mathcal{S_{\sigma}}-\tilde{\tilde{a}}\left(\mathcal{Q}^{-1}\right)^{\lambda\sigma}\mathcal{S^{\mu}\newline_{\sigma\lambda}}\nonumber\\+\frac{2}{3}\tilde{\tilde{c}}\left(\mathcal{R}^{-1}\right)^{\sigma\mu}\mathcal{S_{\sigma}}+\tilde{\tilde{c}}\left(\mathcal{\Re}^{-1}\right)^{\lambda\sigma\mu}\newline_{\rho}\mathcal{S^{\rho}\newline_{\sigma\lambda}}+\tilde{\tilde{C}}\mathcal{F^{\prime}}\left(\mathcal{S}^{-1}\right)^{\mu}=0,\label{122}
\end{eqnarray}
where $\tilde{\tilde{C}} \equiv \frac{\tilde{C}\tilde{c}-\tilde{D}\tilde{a}}{\tilde{a}\tilde{d}-\tilde{b}\tilde{c}}$ and $\tilde{\tilde{D}} \equiv \frac{\tilde{D}\tilde{b}-\tilde{C}\tilde{d}}{\tilde{a}\tilde{d}-\tilde{b}\tilde{c}}$. Thus, by plugging Eqs.(\ref{121}) and (\ref{122}) into Eq.(\ref{120}) we obtain the final form of the gravitational field equations:
 \begin{eqnarray}
\fl(a-1)\nabla_{\sigma}\Big[\left(\mathcal{\Re}^{-1}\right)^{\nu\sigma\mu}\newline_{\rho}\Big]+e\nabla_{\sigma}\Big[\left(\mathcal{\Re}^{-1}\right)^{\mu\sigma\nu}\newline_{\rho}\Big]-\delta^{\nu}_{\rho}
\Big(\frac{2}{3}\overline{a}\left(\mathcal{R}^{-1}\right)^{\sigma\mu}\mathcal{S_{\sigma}}+\frac{2}{3}\overline{b}\left(\mathcal{Q}^{-1}\right)^{\sigma\mu}\mathcal{S_{\sigma}}\nonumber\\ \fl+\overline{b}\left(\mathcal{Q}^{-1}\right)^{\lambda\sigma}\mathcal{S^{\mu}\newline_{\sigma\lambda}}+\overline{a}\left(\mathcal{\Re}^{-1}\right)^{\lambda\sigma\mu}\newline_{\rho}\mathcal{S^{\rho}\newline_{\sigma\lambda}}\Big)-\delta^{\mu}_{\rho}
\Big(\frac{2}{3}\overline{c}\left(\mathcal{R}^{-1}\right)^{\sigma\nu}\mathcal{S_{\sigma}}+\frac{2}{3}\overline{d}\left(\mathcal{Q}^{-1}\right)^{\sigma\nu}\mathcal{S_{\sigma}}\nonumber\\ \fl+\overline{d}\left(\mathcal{Q}^{-1}\right)^{\lambda\sigma}\mathcal{S^{\nu}\newline_{\sigma\lambda}}+\overline{c}\left(\mathcal{\Re}^{-1}\right)^{\lambda\sigma\nu}\newline_{\rho}\mathcal{S^{\rho}\newline_{\sigma\lambda}}\Big)+\frac{2}{3}\left(\mathcal{\Re}^{-1}\right)^{\nu\sigma\mu}\newline_{\rho}\mathcal{S_{\sigma}}-\left(\mathcal{\Re}^{-1}\right)^{\lambda\sigma\mu}\newline_{\rho}\mathcal{S^{\nu}\newline_{\sigma\lambda}}\nonumber\\ \fl -\mathcal{F^{\prime}}\Big[B\left(\mathcal{S}^{-1}\right)^{\nu\mu}\newline_{\rho}+\overline{C}\delta^{\nu}_{\rho}\left(\mathcal{S}^{-1}\right)^{\mu}+\overline{D}\delta^{\mu}_{\rho}\left(\mathcal{S}^{-1}\right)^{\nu}\Big]=0,\label{123}
\end{eqnarray}
where we introduced $\overline{C} \equiv C-c\Big(\frac{\tilde{C}\tilde{c}-\tilde{D}\tilde{a}}{\tilde{a}\tilde{d}-\tilde{b}\tilde{c}}\Big)-b\Big(\frac{\tilde{D}\tilde{b}-\tilde{C}\tilde{d}}{\tilde{a}\tilde{d}-\tilde{b}\tilde{c}}\Big)$ and $\overline{D} \equiv D-d\Big(\frac{\tilde{D}\tilde{b}-\tilde{C}\tilde{d}}{\tilde{a}\tilde{d}-\tilde{b}\tilde{c}}\Big)-b\Big(\frac{\tilde{C}\tilde{c}-\tilde{D}\tilde{a}}{\tilde{a}\tilde{d}-\tilde{b}\tilde{c}}\Big)$. Finally, using Eqs.(\ref{110}) and (\ref{111}) in Eq.(\ref{123}) and then applying the $\mathcal{\Re^{\rho}\newline_{\mu\alpha\nu}}$ product we obtain
\begin{eqnarray}
\fl\frac{\nabla_{\alpha}\left[\sqrt{\vert\mathcal{\Re}\vert}\right]}{\left[\sqrt{\vert\mathcal{\Re}\vert}\right]}=2A\mathcal{S_{\alpha}} - 2\Big(\overline{\overline{a}}\mathcal{R_{\mu\alpha}}\left(\mathcal{\Re}^{-1}\right)^{\lambda\sigma\mu}\newline_{\rho}+2\overline{\overline{b}}\mathcal{R_{\rho\alpha}}\left(\mathcal{Q}^{-1}\right)^{\lambda\sigma} \nonumber\\ +2\overline{\overline{c}}\mathcal{Q_{\mu\alpha}}\left(\mathcal{\Re}^{-1}\right)^{\lambda\sigma\mu}\newline_{\rho} +\overline{\overline{d}}\mathcal{Q_{\rho\alpha}}\left(\mathcal{Q}^{-1}\right)^{\lambda\sigma}\Big)\mathcal{S^{\rho}\newline_{\sigma\lambda}}\nonumber\\ -2\mathcal{F^{\prime}}\Big[\overline{B}\mathcal{\Re^{\rho}\newline_{\mu\alpha\nu}}\left(\mathcal{S}^{-1}\right)^{\nu\mu}\newline_{\rho}-\overline{\overline{C}}\mathcal{R_{\mu\alpha}}\left(\mathcal{S}^{-1}\right)^{\mu}-\overline{\overline{D}}\mathcal{Q_{\nu\alpha}}\left(\mathcal{S}^{-1}\right)^{\nu}\Big]
\end{eqnarray}
with $\overline{B} \equiv \frac{2B}{4-4a-e\Theta}$, $\overline{\overline{C}} \equiv \frac{2\overline{C}}{4-4a-e\Theta}$, and $\overline{\overline{D}} \equiv \frac{2\overline{D}}{4-4a-e\Theta}$. Thus, the last equation leads to the non-local conformal mapping of Riemann curvature as given below:
\begin{eqnarray}
\fl\mathcal{\Re^{\rho}\newline_{\mu\sigma\nu}}\left(x^{\sigma}\right)=\exp\Bigg\{\int_{x_{0}^{\sigma}}^{x^{\sigma}}\Bigg[A\mathcal{S_{\alpha}}+\mathcal{F^{\prime}}\Bigg(\overline{\overline{C}}\mathcal{R_{\rho\alpha}}+\overline{\overline{D}}\mathcal{Q_{\rho\alpha}}\Bigg)\left(\mathcal{S}^{-1}\right)^{\rho}\nonumber \\-\Bigg(\overline{\overline{a}}\mathcal{R_{\mu\alpha}}\left(\mathcal{\Re}^{-1}\right)^{\lambda\sigma\mu}\newline_{\rho}+2\overline{\overline{b}}\mathcal{R_{\rho\alpha}}\left(\mathcal{Q}^{-1}\right)^{\lambda\sigma}+2\overline{\overline{c}}\mathcal{Q_{\mu\alpha}}\left(\mathcal{\Re}^{-1}\right)^{\lambda\sigma\mu}\newline_{\rho}+\overline{\overline{d}}\mathcal{Q_{\rho\alpha}}\left(\mathcal{Q}^{-1}\right)^{\lambda\sigma}\Bigg)\mathcal{S^{\rho}\newline_{\sigma\lambda}}\nonumber\\-\mathcal{F^{\prime}}\Bigg(\overline{B}\mathcal{\Re^{\rho}\newline_{\mu\alpha\nu}}\left(\mathcal{S}^{-1}\right)^{\nu\mu}\newline_{\rho}\Bigg)+\frac{1}{2}\Gamma^{\beta}_{\beta\alpha}\Bigg]dx^{\alpha}\Bigg\}\mathcal{\Re^{\rho}\newline_{\mu\sigma\nu}}\left(x_{0}^{\sigma}\right)
\end{eqnarray}
The torsion determinant included in the scalar function $\mathcal{F^{\prime}}$ now affects the mapping of the Riemann curvature by giving the inverse torsion tensor and the inverse torsion vector to the conformal factor.

\section{Summary}
The main purpose of this study was to examine torsion effects on the non-metrical affine gravity by considering curvature- and torsion-based actions. For each action model, we obtained the torsionful gravitational field equations, and from these equations we found that the Ricci and Riemann curvatures must obey a non-local, exponential rescaling affected by the torsion tensor. In the actions including the Ricci curvature, we also examined the symmetric and antisymmetric Ricci tensors, where the dynamical equations allowed us to construct connection structures such that we are able to obtain the affine connection in the case of a symmetric Ricci tensor and the contracted affine connection in the case of an antisymmetric Ricci tensor. For the action (\ref{7}), we found that the contracted connections obtained in the case of symmetric and antisymmetric Ricci tensors killed the explicit torsion contribution to the rescaling of each kind of Ricci curvature. Moreover, the actions (\ref{7}) and (\ref{61}) resulted in the same form of the rescaling of the symmetric Ricci tensor. In the actions including a torsion determinant in addition to the curvature contribution, we introduced the scalar functions which modified the results of the actions including a purely curvature determinant.
\newline
We want to emphasize that our action models are novel and general enough to reveal salient features of affine gravity. We expect that our work will be important in constructing more realistic models if not for including matter. The models we constructed can be taken further by considering ways of incorporating matter into the affine framework. One of the essential ways for matter coupling to the affine gravity was proposed by making use of the analogy between the affine formalism and canonical mechanics \cite{J1}. In this theory, incorporating matter to the affine gravity can then be provided by applying the covariant Legendre transformations to the affine Lagrangian densities including matter fields. Although these Lagrangian densities do not depend on any metric, the fields can be equipped with a metric tensor due to the canonical structure of this affine formulation, where the metric and the connection of that metric play the roles of the momentum and the generalized coordinate. Another way was based on the construction of the metric tensor only from the affine connection through the symmetric, contracted square of the torsion tensor \cite{P3}. Proposing a Lagrangian density composed from this torsional metric tensor together with the curvature it can be shown that if the matter fields couple to the torsion square through the metric tensor one may then obtain the Einstein field equations with its matter part as well as the cosmological constant. However, how to incorporate matter into the affine theory of gravity in our models is not trivial and has not been developed yet. But following the same techniques as used in \cite{J1, P3}, that is, by extending our models with the  Lagrangian densities involving matter fields or proposing a torsion dependent metric, we leave this construction to near future detailed work, and in this case, one may clearly see shifts of these models from the general relativity when applied to the Solar System so that possible explanations of some phenomena such as dark matter or the avoidance of singularities will be explored.
\\
In addition, there has been a recent claim in Ref. \cite{Demir:2015kxa} that the naturalness problem in the standard model may be dealt with by the Eddington formulation of gravity. This idea also addresses the relevancy of the affine formulation of gravity to the standard model to clarify some fundamental problems in it. In this respect, our purely affine actions with their results may also be useful for and applied to the standard model in the future within the claim in \cite{Demir:2015kxa}.
\\
In summary, it is crucial and necessary to properly study and develop the affine theory of gravity due to its more fundamental nature and generality than general relativity. Although it is not trivial and has not been studied yet, via the improvement of the purely affine theory we expect important applications and contributions of our results to the standard model of particle physics, cosmology, astroparticle physics, black holes, neutron stars, supernovas, and other physical areas, where the gravitational field is very strong, to clarify some fundamental problems emerging from them.

\section*{Acknowledgements}
I am very thankful to Durmu\c{s} Ali Demir and  Hemza Azri for fruitful suggestions and discussions on various aspects of the present work. I am also grateful to \.{I}smail Hakk{\i} Duru for his support and much valuable help.

\end{document}